\documentclass[twocolumn,prb,showpacs,multicol,amsmath,amssymb]{revtex4}
\usepackage[dvips]{graphicx}
\usepackage{graphicx}
\usepackage{dcolumn}
\usepackage{bm}
\usepackage{graphics}
\usepackage{soul}
\usepackage{epsfig,color}

\newcommand{\be}{\begin{equation}}
\newcommand{\ee}{\end{equation}}
\newcommand{\bea}{\begin{eqnarray}}
\newcommand{\eea}{\end{eqnarray}}

\begin{document}
\title[J. Vahedi ]{Effects of quantum interference on the electron transport in the semiconductor$/$benzene$/$semiconductor junction}
\author{Javad Vahedi$\footnote{email: javahedi@gmail.com\\ Tel: (+98) 911-1554504\\Fax: (+98) 11-33251506}$, Zahra Sartipi}
\address{ Department of Physics, Sari Branch, Islamic Azad University, Sari, Iran.}
\date{\today}
\begin{abstract}
Using the tight-binding model and the generalized Green's function formalism, the effect of quantum interference on the electron transport through the benzene molecule in a semiconductor/benzene/semiconductor junction is numerically investigated. We show how the quantum interference sources, different contact positions and local gate, can control the transmission characteristics of the electrode/molecule/electrode junction. We also study the occurrence of anti-resonant states in the transmission probability function using a simple graphical scheme (introduced in Ref.[Phys. Chem. Chem. Phys, 2011, 13, 1431]) for different geometries of the contacts between the benzene molecule and semiconductor(silicon and titanium dioxide) electrodes.
\end{abstract}

\maketitle

\section{INTRODUCTION}\label{sec1}
Recently, harnessing quantum interference (QI) effect as a versatile tool  to design of single molecule devices as well as controlling the current through it has attracted many attention\cite{a2,a3,a4,a5,a6,a7,a8}. QI  hapens when electron waves go through a meso/nano junctions phase-coherently. One of the important factors to study electron transport through the mesoscopic junctions is to control the quantum interference effect of electron wave related to the symetry that device adopts within the junctions. Regarding the employment of QI as an enabling tool for the execution of molecular switches, logic gates,  data strong elements and thermoelectronic devices in single molecule devices, two different proposals  have been suggested. One way to induce QI in molecular junction is to use a local gate potential to tune the position of induced transmission nodes relative to bias window, and the other is to control the electron transmission through chemical/conformational modification of side groups to aromatic molecules\cite{a3,a4,a5}.
\par
 Studying wave-guides for semiconductor nano-structures was the first ground for signaturing QI, though it has been reported early on theoretically\cite{a13} and experimentally\cite{a14,a15} that electron transport through a benzene molecule connected in a meta-configuration is strongly diminished in comparing with two ortho and para configuration between the molecule and the leads. This connection of the conductance in  the benzene configuration has been recognized as QI and has been described in terms of phase shifts of transmission channels, and it has also been observed  through another generic aromatic molecules.
\par
On the other hand, some theoretical and experimental efforts, have mainly focused on systems with two metal  leads\cite{a15,a16,a17,a18,a19,a20,a21,a22,a23,a24} often gold, intriguing physics, occurs when one or both electrodes are replaced with semiconductors\cite{a24,a25,a26,a27,a28,a30,a31,a32,a322},including negative differential resistance and rectification\cite{a33}. Indeed, semiconductor/molecule/semiconductor hybrid junctions with self-assembled mono-layers (SAM) are considered as promising candidates to create functional devices for molecular electronics. Semiconductor substrates are widely used in many applications. Multiple practical uses involving these materials require the ability to tune their physical (bandgap, electron mobility) and chemical (functionalization, passivation) properties to adjust those to a specific application. In the work reported in the present article we investigated the effect of quantum interference on the electronic conductance of semiconductor/benzene/semiconductor junction which no studies have been reported to date. Using tight-binding model and a generalized Green's function method in the Landauer-B\"{u}ttiker formalism, we have tackled this problem and the variation of interference condition determined by replacement three different configurations of the benzene molecule (ortho, meta and para) connected to the semiconductor/metal electrodes.
\begin{figure}
 \includegraphics[width=.3\columnwidth]{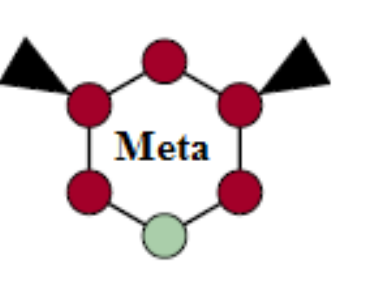}
 \includegraphics[width=.3\columnwidth]{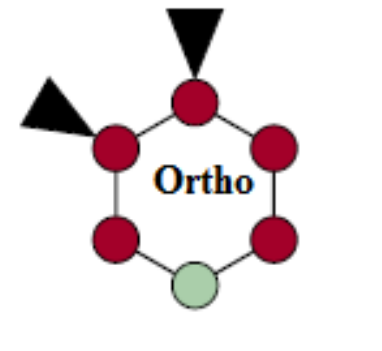}
 \includegraphics[width=.3\columnwidth]{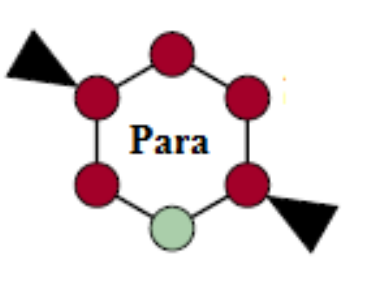}
 \caption{Schematic representation of three meta, ortho and para configurations. Local gate is shown in light green}
\label{fig1}
\end{figure}

\par
\begin{figure}
 \includegraphics[width=.8\columnwidth]{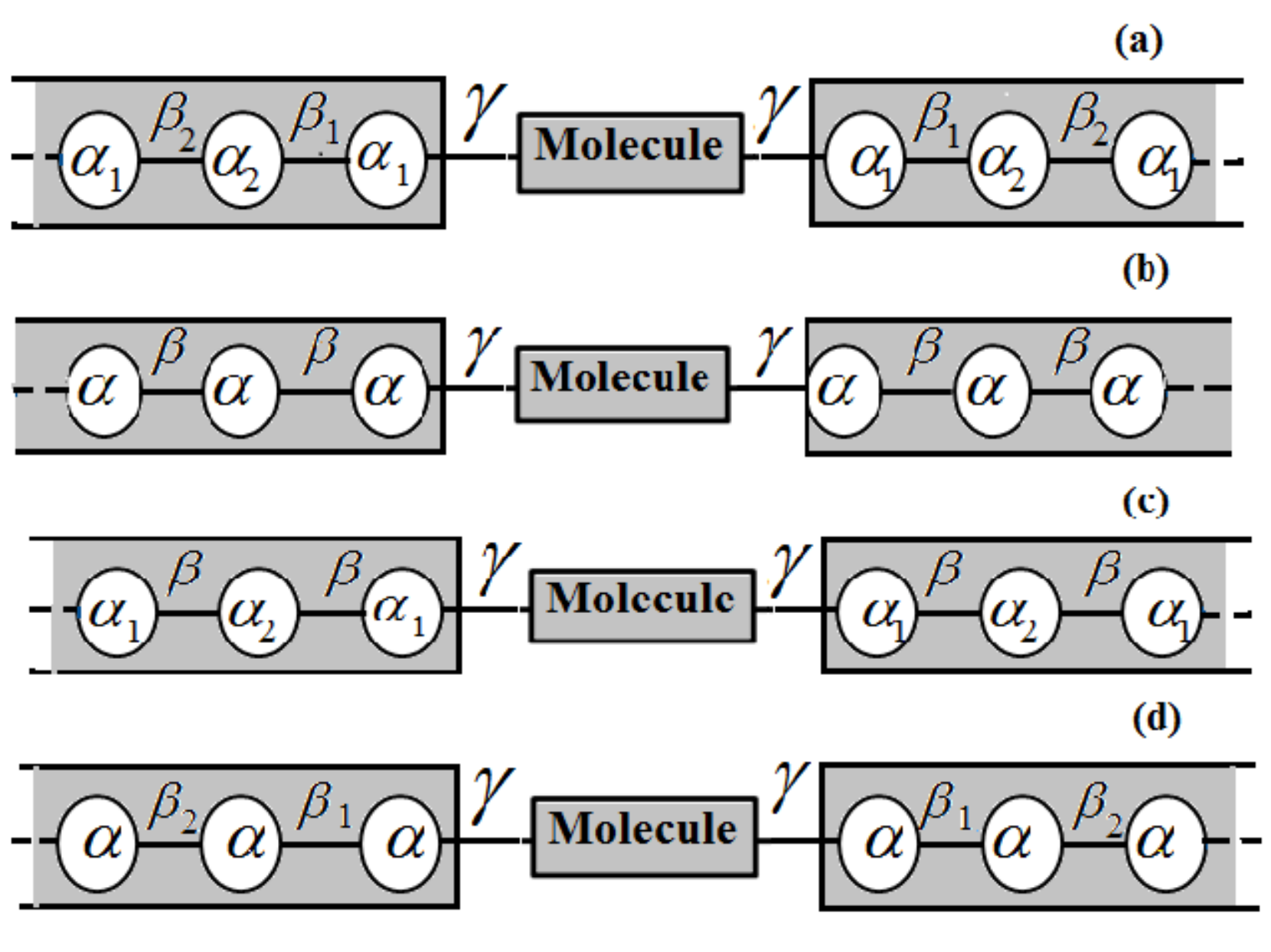}
 \caption{Schematic representation of (a) Koutecky-Davison (KD), (b) Newns-Anderson (NA), (c) Alternative site(AS) and (d) Alternative bond (AB) models.}
\label{fig2}
\end{figure}

Semiconductor modeling is acquirable via some perturbations of the Newns-Anderson (NA) model as seen in Fig.(\ref{fig2}-b). One such model, introduced by Koutecky and Davison (KD) is the generalization of the simple tight binding model, which alternates both the site energies $\alpha_1, \alpha_2$ and intersite couplings $\beta_1, \beta_2$ with nearest-neighbor in tight-binding picture Fig.(\ref{fig2}-a). This model may consider each unit cell as an atom, with two sites states corresponding to $s$ and $p$ orbitals. The coupling between $s$ and $p$ orbitals in the same unit cell describes their overlap, and the coupling across unit cells describes the interatomic bonding. Neighboring $s-s$ and $p-p$ interactions are neglected in the simple tight-binding model. The KD model has three important limits. The first one can achieve in the combined limit $\alpha\rightarrow 0, \beta_1\rightarrow \beta_2\equiv\beta$ which produces the Newns-Anderson model as seen in Fig.(\ref{fig2}-b) that $\alpha$ and $\beta$ are the on-site energy and intersite hopping energies, respectively, in tight-binding model\cite{a33}. The second one is the limit $\alpha_1=-\alpha_2, \beta_1\rightarrow \beta_2\equiv\beta$. This alternating site (AS) model has been used to describe titanium dioxide where the site energies ($\alpha$ and $-\alpha$) correspond to the different atoms Fig.(\ref{fig2}-c)\cite{a33}. The third one is the limit  where the model alternates bonds(AB)Fig.(\ref{fig2}-d). The AB model has been used to model silicon and germanium, where the bond disparities($\beta_1$ and $\beta_2$) are related to the orbital hybridization\cite{a33}.
\par
Intercepting crystal into two  noninteracting segments causes symmetry breaking and surface states arising. The surfaces can exhibit dangling bonds, potentially leading to reconstructions surface states, with densities localized near the surface\cite{a33,a34,a35}. Three $0, 1$ and $2$ surface states are created in both AS and AB models. For AB model $0, 2$ and $1$ surface states are defined as $|\beta_1|>|\beta_2|$ in both leads, $|\beta_1|<|\beta_2|$ in both leads and $|\beta_1|>|\beta_2|$($|\beta_1|<|\beta_2|$) in source(drain), respectively. For AS model $0, 2$ and $1$ surface states are defined as $\alpha_1<0$ in both leads, $\alpha_1>0$ in both leads and $\alpha_1>0$ ($\alpha_1<0$) in source(drain), respectively.
\par
The paper is organized as follows. In section (II), we summarize the model, Hamiltonian and formalism. More details on the (KD) model are given in the appendix A. In section (III), we present our results for QI in the benzene molecule by numerical calculations and graphical rules. Finally, conclusion is given in section IV.
\\
\section{Computational Scheme} \label{sec2}
We start with Fig.(\ref{fig1}),where a benzene molecule which subject to a local gate potential is attached to semi-infinite semiconductor electrode. We have also considered metal electrodes, for comparison. Leads are characterized by electrochemical potentials and respectively, under the non-equilibrium condition when an external bias voltage is applied. Both leads have almost the same cross section as the sample to reduce the effect of the scattering induced by a wide-to-narrow  geometry of the sample-lead interface. The whole system is described within a single electron picture by a tight-binding Hamiltonian with nearest-neighbor hopping approximation. The Hamiltonian representing the entire system can be written as a sum of four terms
\begin{equation}
\emph{H}=H_L+H_R+H_M+H_C
\label{e1}
\end{equation}
where $H_{L,(R)}$ represents the Hamiltonian of the left (right) electrode which is described within the tight-binding approximation, $H_{M}$ is the Hamiltonian of the molecule and the last term in Eq.(\ref{e1}) is the Hamiltonian of coupling of electrodes to the molecule. The Hamiltonian of KD  model as a semiconductor lead in the tight-binding picture is expressed as follows
 \begin{widetext}
\begin{equation}
\emph{H}_{KD}=\alpha\sum_{j=1}^N\big(\chi_{2j-1}^\dag\chi_{2j-1}-\chi_{2j}^\dag\chi_{2j}\big)+\Big[\beta_1\sum_{j=1}^{N}\big(\chi_{2j-1}^\dag\chi_{2j}\big)
+\beta_2\sum_{j=1}^{N-1}\big(\chi_{2j+1}^\dag\chi_{2j}\big)+h.c\Big]
\label{e2}
\end{equation}
\end{widetext}
Here $\chi_{j}^\dag$ ($\chi_{j}$) denotes the creation (annihilation) operator of an electron at site $j$. Besides, its three Hamiltonian limits NA, AS and AB models can be expressed in the following forms,

\begin{eqnarray}
\emph{H}_{NA}&=&\beta\sum_{j=1}^N \chi_{j}^\dag\chi_{j}+h.c\nonumber\\
\emph{H}_{AS}&=&\alpha\sum_{j=1}^N\big(\chi_{2j-1}^\dag\chi_{2j-1}-\chi_{2j}^\dag\chi_{2j}\big)\nonumber\\
             &+&\beta\sum_{j=1}^{2N-1}\chi_{j}^\dag\chi_{j}+h.c\nonumber\\
\emph{H}_{AB}&=&\beta_1\sum_{j=1}^N\chi_{2j-1}^\dag\chi_{2j}-\beta_2\sum_{j=1}^{N-1}\chi_{2j+1}^\dag\chi_{2j}+h.c
\label{e3}
\end{eqnarray}

 the Hamiltonian of benzene can be described as follows:
 \begin{equation}
\emph{H}_M=\sum_i\epsilon_iC_i^\dag C_i+t\sum_{<ij>}\big( C_i^\dag C_j+C_j^\dag C_i\big),
\label{e4}
\end{equation}
Here$C_i^\dag(C_i)$ denotes the creation (annihilation) operator of an electron at site $i$ in the benzene molecule. $\epsilon_i$  and $t$ are the on-site energy and nearest-neighbor hopping integral in the benzene molecule, respectively. The last term in Eq.(\ref{e1}) is the coupling part and given by
 \begin{equation}
\emph{H}_C=\sum_i\gamma_{c(j,i)}\big(\chi_i^\dag C_i+h.c)\big),
\label{e5}
\end{equation}
where the matrix elements$\gamma_{c(j,i)}$ represents the coupling strength between the molecule and electrodes. The Green's function of the system can be written as
 \begin{equation}
\emph{G(E)}_C=\frac{1}{E\textbf{1}-\emph{H}_M-\Sigma_L(E)-\Sigma_R(E)},
\label{e6}
\end{equation}
where $\textbf{1}$ stands for the identity matrix and $\Sigma_L(E)(\Sigma_R(E))$ is the self-energy matrix resulting from the coupling of the benzene molecule to the left (right) electrode and given by
 \begin{eqnarray}
\Sigma_{L(R)}(E)&=&\gamma_c^\ast g_{L(R)}\gamma_c\nonumber\\
                &=&\Lambda_{L(R)}(E)-\frac{i}{2}\Gamma_{L(R)}(E),
\label{e7}
\end{eqnarray}
where $\gamma_c$ is the coupling matrix and will be non-zero only for the adjacent points in the benzene molecule and electrode. In Eq.(\ref{e7}), $\Lambda_{L(R)}$ is the real part of self-energy which correspond to the shift of energy levels of the benzene molecule and the imaginary part $\Gamma_{L(R)}(E)$ of the self-energy represents the broadening of these energy levels. In the follow shifts and broadening  terms of energy levels are given for KD model\cite{a33}.
\begin{widetext}
\begin{eqnarray}
\Gamma_{KD}(E)&=&\frac{\gamma^2}{\beta_2^2}\sqrt{\frac{\big[\alpha^2+(\beta_1+\beta_2)^2-E^2\big]\big[E^2-\alpha^2-(\beta_1-\beta_2)^2\big]}{(E-\alpha)^2}}\nonumber\\
 \frac{\Lambda_{KD}(E)}{\gamma^2}&=&\frac{E^2-\alpha^2-\beta^2_1+\beta^2_2+\Theta_{KD}(E)\sqrt{\big[E^2-\alpha^2-(\beta_1-\beta_2)^2\big] \big[E^2-\alpha^2-(\beta_1+\beta_2)^2\big] }}{2\beta_2^2(E-\alpha)}\nonumber\\
 \Theta_{KD}(E)&=&\Theta(\alpha^2+(\beta_1-\beta_2)^2-E^2)-\Theta(E^2-\alpha^2-(\beta_1+\beta_2)^2),
\label{e8}
\end{eqnarray}
\end{widetext}
where $\Theta$ is the sign function. Having $\Gamma_{KD}, \Lambda_{KD}$, one can easily find the other models relations. By applying $\alpha\rightarrow0$,$\beta_1\rightarrow\beta_2\equiv\beta$ we have $\Gamma_{KD}, \Lambda_{KD}\rightarrow\Gamma_{NA}, \Lambda_{NA}$. Also taking the limit $\beta_1\rightarrow\beta_2\equiv\beta$  and $\alpha\rightarrow0$ gives relations for the AS and AB models,respectively. In Eq.(\ref{e7}), $g_{L(R)}$ correspond to the Green's functions for the left(right) electrode. The Green's functions for the KD model can be written as follows\cite{a33}:
\begin{widetext}
\begin{eqnarray}
g_{KD}&=&\frac{E^2-\alpha^2+\beta_1^2+\beta_2^2\pm \sqrt{\big[E^2-\alpha^2-(\beta_1-\beta_2)^2\big] \big[E^2-\alpha^2-(\beta_1+\beta_2)^2\big]}}{2\beta_2^2(E-\alpha)}
\label{e9}
\end{eqnarray}
\end{widetext}
Some of the details on the derivation are further explained in Appendix A.
\par
The transmission probability between reservoirs is given as
\begin{equation}
T(E)=Tr\big( \Gamma_LG^r\Gamma_RG^a\big),
\label{e10}
\end{equation}
where $G^r$ and $G^a$ are the retarded and advanced Green's function, respectively. The broadening matrix is defined as the imaginary part of the self-energy $\Gamma_{L(R)}=-2 Im(\Sigma_{L(R)})$. Transmission function tells us the rate at which electrons transmit from the left to the right electrode by propagating through the molecule. Relying on Landauer-B\"{u}ttiker formula, we evaluate the current as a function of the applied bias voltage
\begin{equation}
I(V)=\frac{2e}{h}\int_{-\infty}^\infty dET(E,V)\big[f_L(E,V)-f_R(E,V)\big]
\label{e11}
\end{equation}
where $f_{L(R)}$ is the Fermi distribution function at the left (right) electrode with chemical potential $\mu_{L(R)}=E_F\pm\frac{V}{2}$ and Fermi energy $E_F$.
\section{Numerical Results} \label{sec3}
Here, we present results of the numerical calculation based on the formalism described in section\ref{sec2}. We have calculated the transmission through the benzene molecule for ortho, meta and para configurations. For simplicity, we chose the on-site energy equal to zero, i.e., $\varepsilon=0$ without loss of generality. Furthermore, the value of the hopping energy is set as $t=-2 eV$ in the tight-binding picture. Local gate potential  is applied on site $5$ with$(\varepsilon_g=0.5eV)$ (see Fig.\ref{fig1}, local gate is shown in light green spot). The NA, AB and AS model parameters have listed in Table-\ref{table1}\cite{a33}. As a reference energy, the Fermi energy of  electrodes is set $E_F = 0$. The temperature is also set as $T=4K$.
 \begin{table}[b]
\caption{\label{table1}%
Model Parameters for Au, Si, and TiO$_2$}
\begin{ruledtabular}
\begin{tabular}{ccdddd}
material&model&
\multicolumn{1}{c}{\textrm{$|\alpha|(eV)$}}&
\multicolumn{1}{c}{\textrm{$\beta_1(eV)$}}&
\multicolumn{1}{c}{\textrm{$\beta_2(eV)$}}&
\multicolumn{1}{c}{\textrm{$\gamma(eV)$}}\\
\hline
Au\cite{a11}&NA& &-8.95& & -0.45\\
Si&AB&  & -1.60 & -2.185 & -1.0 \\
TiO$_2$ &AS & 1.6 & -2 &  & -1.0 \\
\end{tabular}
\end{ruledtabular}
\end{table}

\subsection{Metal electrode}
\begin{figure}
 \includegraphics[width=0.9\columnwidth]{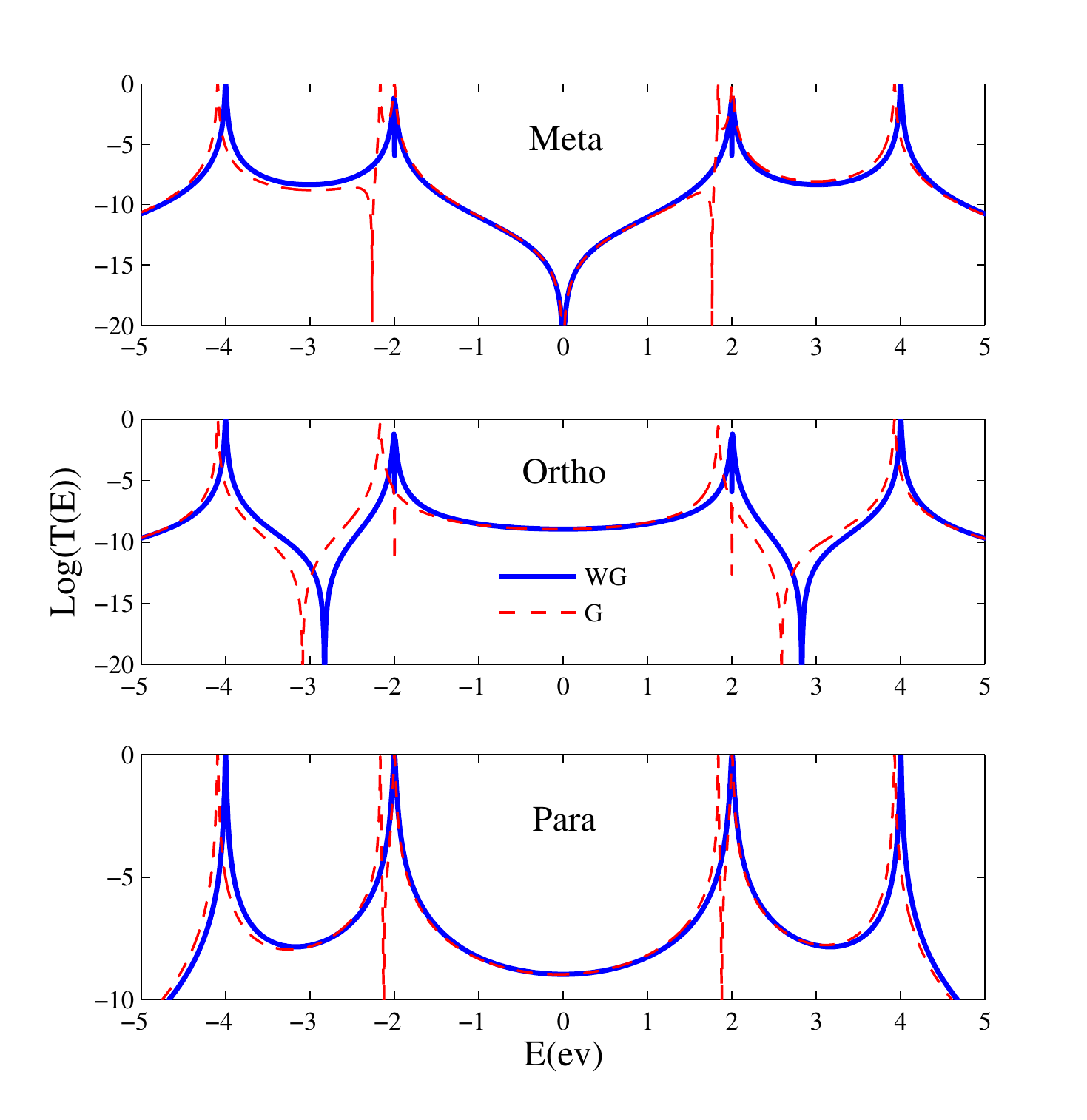}
 \includegraphics[width=0.9\columnwidth]{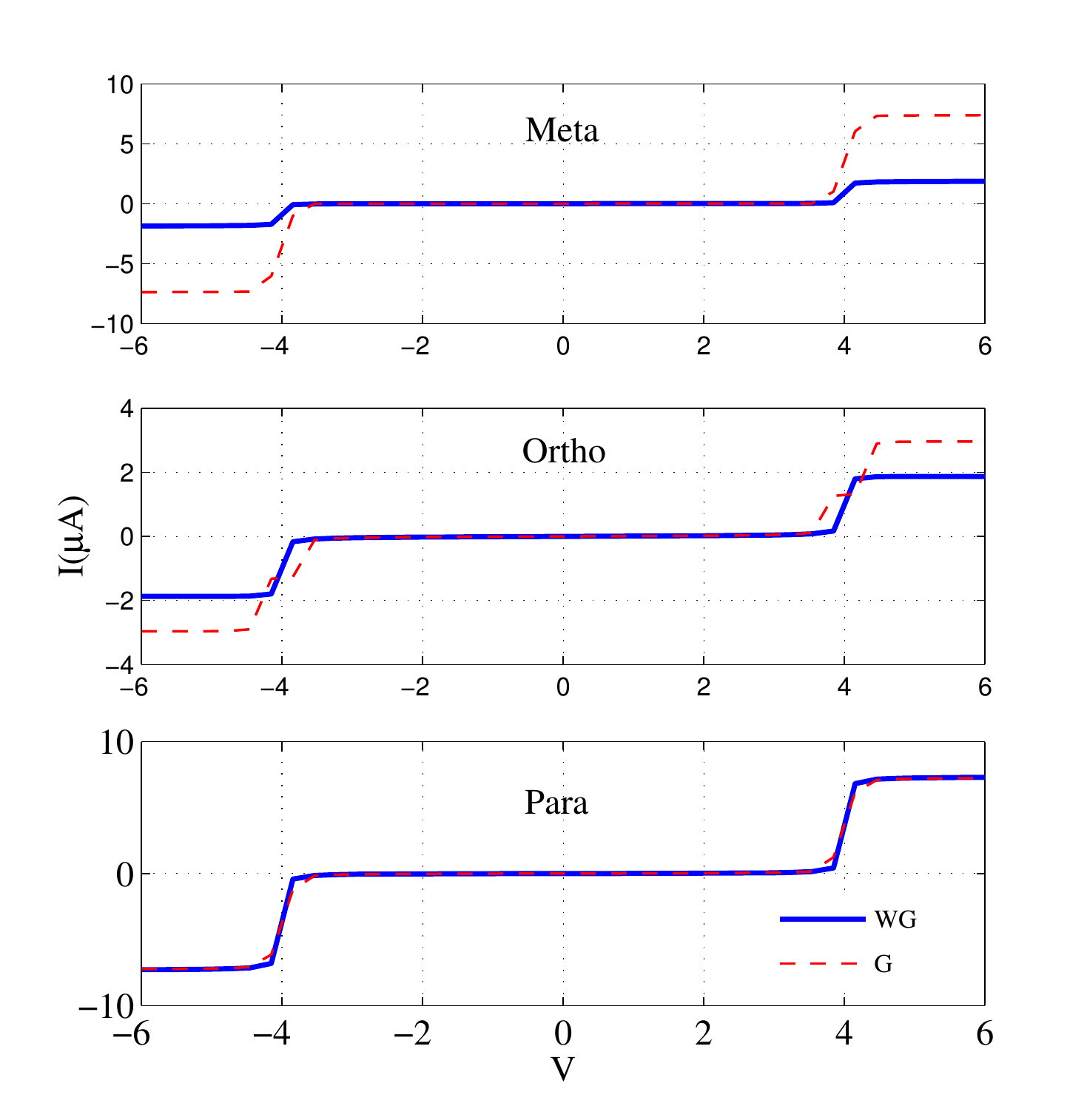}
 \caption{(Color online) The top panel is the logarithmic  scale of  transmission function versus energy of the gold/benzene/gold junction  for meta, ortho and
  para configurations. The bottom panel is the current-voltage (I−V) for meta,  ortho and para configurations. The dotted(solid)line corresponding to the gated (ungated) potential.} \label{fig3}
\end{figure}

We have plotted the logarithmic scale of transmission function versus energy of the gold/benzene/gold junction for meta, ortho and para configurations (see the top panel of Fig.(\ref{fig3}). The dotted lines are transmission function with a gate (G) potential applied on site $5$ and the solid lines are the transmission through ungated (WG) benzene for comparison. For an electron with energy $E$, that comes from the left connection, the probability  of transmission function  reaches  its saturated value (resonance peaks)  for the specific energy values. These  resonance peaks  are related to the eigenenergies of the individual benzene. When the electron travel  from  the left electrode to the right one through the benzene molecule, the electron waves propagating along the two branches of benzene may suffer a relative phase shift. Consequently, there might be constructive or destructive interference due to the superposition  of the  electronic wave functions along the various  pathways. Therefor, the transmission probability will change and some anti-resonance behaviour may arise. In the top panel of Fig.(\ref{fig3}), for meta and ortho configurations one can easily see such anti-resonance behaviours (solid lines). These anti-resonance  states are related to the quantum destructive interference effect. In contrast, for para configuration in the absence of the gate potential there is not such an anti-resonance behaviour
which is referred to the constructive superposition of the electron waves.
\par
In what follows, we will apply a local gate potential and check its effects on the transport properties of three configurations. To this end, we exploit a simple but versatile graphical method proposed recently\cite{a36}. The authors showed that using the graphical method one can easily detect the presence/absence of QI induced transmission nodes only from the topological structure of molecules. Here, we point out some essential features and rules of graphical method. A detailed description can be found in Ref. [36]. With some mathematical task one can recast the Eq.(\ref{e10}) as follows
\begin{equation}
      T(E)=\gamma(E)^2 |G_{1N}(E)|^2
      \label{e12}
\end{equation}
where, $G_{1N}(E)$ is the $(1N)$ matrix element of the Green function, and $\gamma(E)=[\Gamma_L(E)]_{11}=[\Gamma_R(E)]_{NN}$. Often we can neglect the energy dependence on the lead coupling strength, both for metal and semiconductor electrodes then QI effects are included in $G_{1N}(E)$. Using Cramer's rule the matrix element $ G_{1N}(E)$ is defined
\begin{equation}
G_{1N}(E)=\frac{C_{1N} (EI-H_ {M})}{det(EI-H_ {M}-\Sigma_l-\Sigma_R)}
\label{e13}
\end{equation}
where $ C_{1N}(EI-H_ {M}) $ is the $(1N)$ co-factor of $(EI-H_ {M})$ determined as the determinant of the matrix obtained by removing the first row and the N-th column from $(EI-H_ {M}-\Sigma_l-\Sigma_R) $ and multiplying it by $ (-1)^{1+N} $. Since we consider solely sites $1$ and $N$ pair to the left and right electrodes, the
elimination of the first row and Nth column entirely removes $ \Sigma_{L,R} $ in the co-factor. Therefore we will concentrate on $C_{1N}(EI-H_ {M})$ and represent the determinant graphically. The transmission zeros can be defined from the zeros of the co-factor (condition for complete destructive interference).
\begin{equation}
(-1)^ {N+1}C_{1N}(EI-H_ {M})=0
\label{e14}
\end{equation}
to understand how the graphical rules follow from Eq.(\ref{e14}), we use the Laplace's formula for evaluating of co-factor. $det(A)=\sum A_{ij}(-1)^{i+j} M_{ij}$, where $M_{ij}$ is the determinant obtain from removing first row and N-th column of the matrix (A). So the elements of co-factor gives an equation for the transmission nodes in  certain energies.
\par
We start with Hamiltonian of benzene molecule.
\begin{equation}
       H_{M}=
      \left[ {\begin{array}{cccccc}
      \varepsilon_1 & t_{12} &0 & 0 & 0&t_{16}  \\
      t_{21} & \varepsilon_2 &t_{23} & 0 & 0&0  \\
       0 & t_{32} & \varepsilon_3 & t_{34} & 0&0 \\
       0 & 0 & t_{43} & \varepsilon_4 & t_{45}&0  \\
       0 & 0 & 0 & t_{54}&\varepsilon_5 & t_{56}\\
       t_{61} & 0 & 0 & 0& t_{65}&\varepsilon_6\\
      \end{array} } \right]
       \label{e16}
\end{equation}
Note that for different configuration the Hamiltonian will be unchanged and we have just different co-factor for different meta, ortho and para configurations. On-site energies are set zero except site $5$ which we put a local gate $\varepsilon_5=\varepsilon_g$, hopping integrals are also set equal as $t$. So it is simple to adjust co-factor for different meta, ortho and para configurations. Co-factor in para configuration is achievable solely by omitting 1st row and 4th column of matrix $C_{14}(EI-H_ {M})$
  \begin{equation}
       C_{14}(E-H_{M})=
      \left| {\begin{array}{ccccc}
      -t & E & -t & 0 & 0  \\
       0 & -t & E & 0 & 0  \\
       0 & 0 & -t & -t & 0  \\
       0 & 0 & 0 & E-\varepsilon_g & -t\\
       -t & 0 & 0 & -t & E\\
      \end{array} } \right|
       \label{e17}
    \end{equation}
when we write down the elements of co-factor, we reach an equation which determines the transmission nodes energy (transmission zeroes). In general its a tedious work and graphical method circumvents this complexity.
\begin{figure}
\includegraphics[width=0.9\columnwidth]{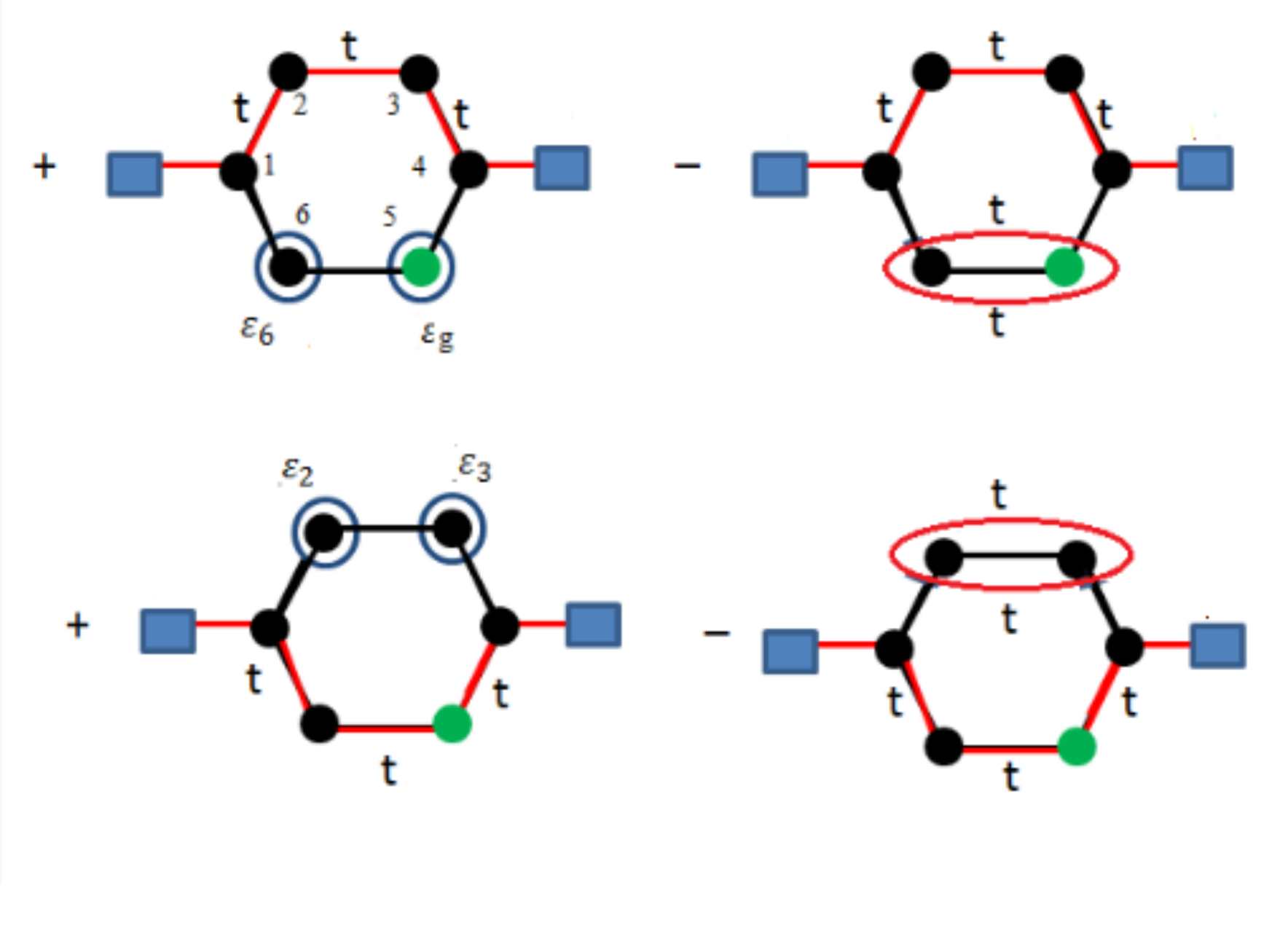}
\caption{(Color online) All graphical diagrams determining the transmission zeros. Six sites connected to leads at site $1$ and $4$ in para configuration.}
\label{fig4}
\end{figure}
\begin{figure}
\includegraphics[width=1.0\columnwidth]{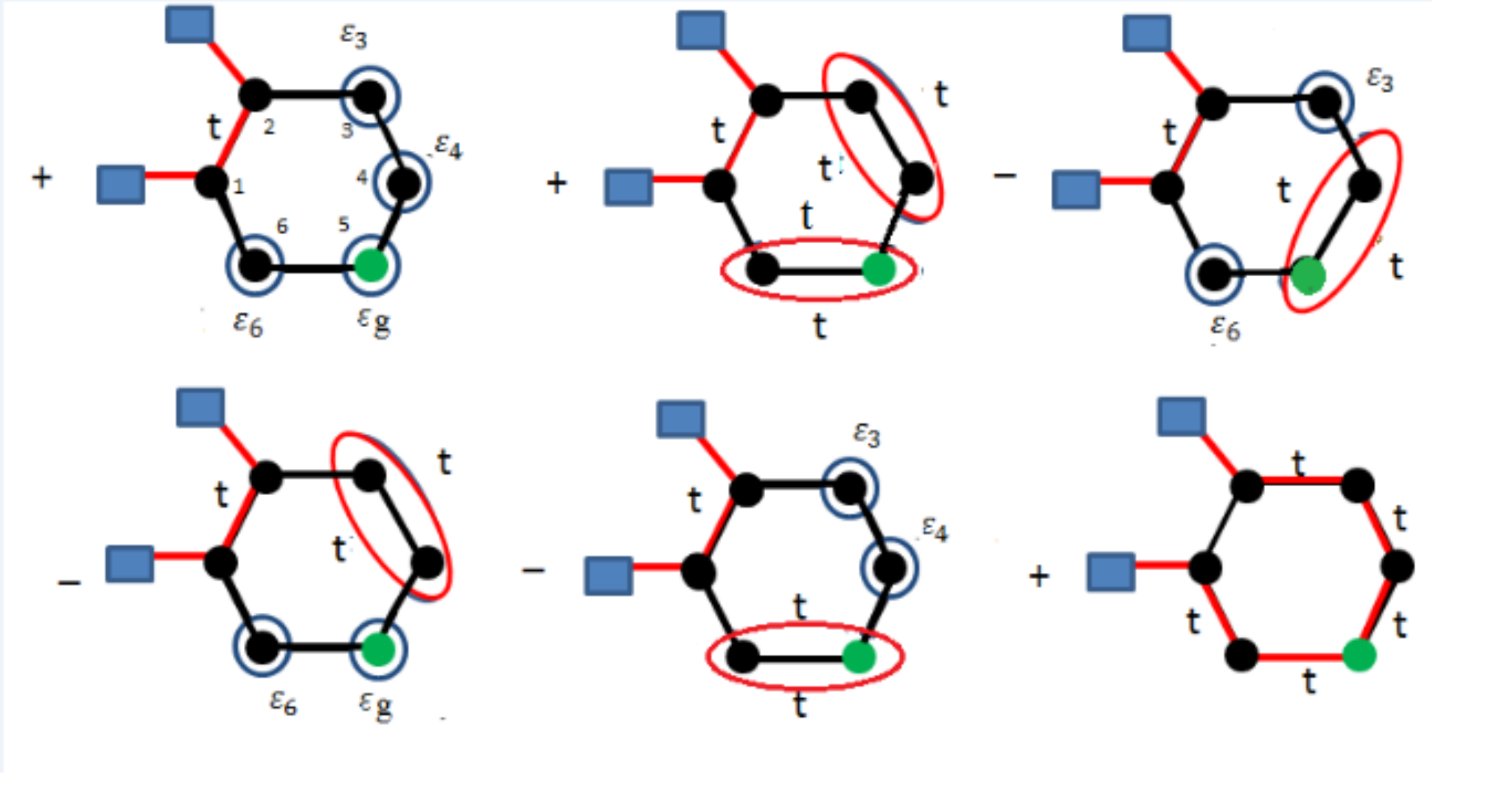}
\caption{(Color online)All graphical diagrams determining the transmission zeros.  Benzene molecule connected to leads at site $1$ and $2$ in ortho configuration.}
\label{fig5}
\end{figure}
\begin{figure}
\includegraphics[width=0.85\columnwidth]{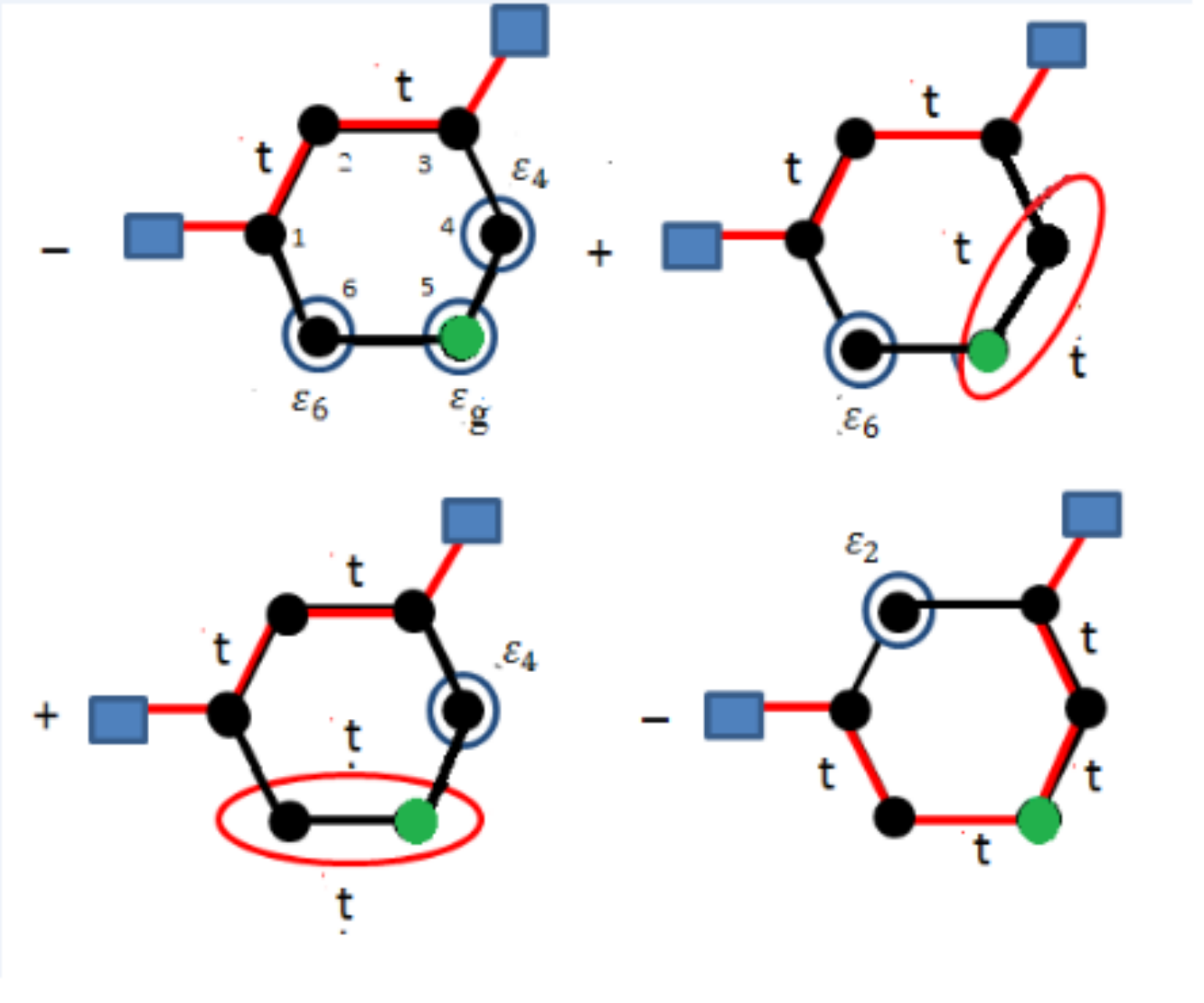}
\caption{(Color online) All graphical diagrams determining the transmission zeros.  Benzene molecule connected to leads at site $1$ and $3$ in meta configuration.}
\label{fig6}
\end{figure}
\par
In  Fig.(\ref{fig4}) we have depicted  all the diagrams based on graphical rules that display how the graphical method is applied to predict the possible existence of QI in an energy range benzene connected to leads in para formation. The four terms in the co-factor can be arisen graphically according to the following rules: (i) in each diagram between the external sites $i$ and $j$, there is only one continuous path connecting them. (ii) at all remaining sites, internal sites (sites other than $i$ and $j$) there is one outgoing and one incoming path or either have an on-site loop. ( iii) each diagram has a sign that is $(-1)^p$ where p is the total number of closed hopping loops and on-site loops. (iv) between two sites $i$ and $j$, only one path can be drawn if they have none-zero hopping elements, $t_{ij}$.  By converting these diagrams into a polynomial, the co-factor $C_{14}(E-H_M)$ is in hand. Now we write down polynomial function for para configuration based on above conditions as follows
\begin{eqnarray}
&&\{+t^3(\varepsilon_g-E)(\varepsilon_6-E)-t^5-t^5\nonumber\\
&&+t^3(\varepsilon_3-E)(\varepsilon_2-E)\}=0\nonumber\\
&\Rightarrow& E_1= 1.87 , E_2= -2.12
\end{eqnarray}%
 where $E_1$ and $E_2$ are two points in which transmission spectrum vanishes. The derivation of the co-factor $C_{1N}(E-H_{M})$ for the benzene molecule in ortho and meta configurations will follow the same methodology of the co-factor by the graphical scheme in para configuration. We have drawn all the generalized diagrams determining the transmission zeroes, for ortho and meta configurations of benzene molecule in Fig.(\ref{fig5}))and Fig.(\ref{fig6})), respectively.
 so for the ortho configuration
\begin{eqnarray}
&&\{+t(\varepsilon_3-E)(\varepsilon_4-E)(\varepsilon_6-E)(\varepsilon_g-E)+t^5+t^5\nonumber\\
&&-t^3(\varepsilon_g-E)(\varepsilon_6-E)-t^3(\varepsilon_4-E)(\varepsilon_3-E)\nonumber\\
&&-t^3(\varepsilon_3-E)(\varepsilon_6-E)\}=0\nonumber\\
&\Rightarrow& E_1=2  ,E_2=-2  ,E_3\simeq-3.09  ,E_4\simeq2.59
\end{eqnarray}
and for the meta configuration
\begin{eqnarray}
&&\{-t^2(\varepsilon_4-E)(\varepsilon_6-E)(\varepsilon_g-E)+t^4(\varepsilon_4-E)\nonumber\\
&&+t^4(\varepsilon_6-E)-t^4(\varepsilon_2-E)\}=0\nonumber\\
&\Rightarrow& E_1=1.76 , E_2=-2.26 ,E_3=0
\end{eqnarray}
\par
Comparing the above result shows a good agreement with our numerical results based tight-binding method in Fig.(\ref{fig3}). More precisely, the transmission zeros points are exactly corresponding to roots of above equations for all three possible configurations of the benzene molecule.
\par
In order to provide a deep understanding  of the electron transport, we have depicted the current-voltage for three ortho, meta and para configurations in the bottom panel of Fig.(\ref{fig3}). An applied voltage shifts the chemical potentials of two electrodes relative to each other by $eV$, with $e$ the electronic charge. When the benzene level is positioned  within such bias window, current will flow. With consideration to the current profile, it shows staircase-like structure  with fine steps versus of  the applied bias voltage. This is due to the existence of the sharp resonance peaks in the transmission  spectrum, since the current is computed  by the integration of the transmission function Eq.(\ref{e11}). With the increase of the bias voltage the electrochemical potentials on the electrodes are shifted gradually, and finally cross one of the quantized energy levels of the benzene. Therefore, a current channel is opened up which provides a jump in the characteristic.  For the same voltage the current magnitude of the three ortho, meta and para configurations exhibit different responses to the applied
bias voltage. This is due to the quantum interference effects of the electron waves traversing through the different branches of the benzene molecule.
By comparing the current plots in the presence and absence of the gate potential, two features can be readily found. First, three configurations ortho, meta and para show different current magnitudes and in this regard meta configuration shows profound responses to the gate potential. The second one is different threshold voltage. Indeed, the presence of gate potential indicates smaller threshold voltage than the ungated one.
\par
Now, we proceed to investigate the effects of  semiconductor electrodes  on electronic transport. Note that we neglect all the interactions between current carriers and electron  correlations are associated only with the Pauli principle. Then we can control the current function the electrode/benzene/electrode  system by contacting the benzene to the electrodes from different locations as a geometrical interference.
\begin{figure}
 \includegraphics[width=0.8\columnwidth]{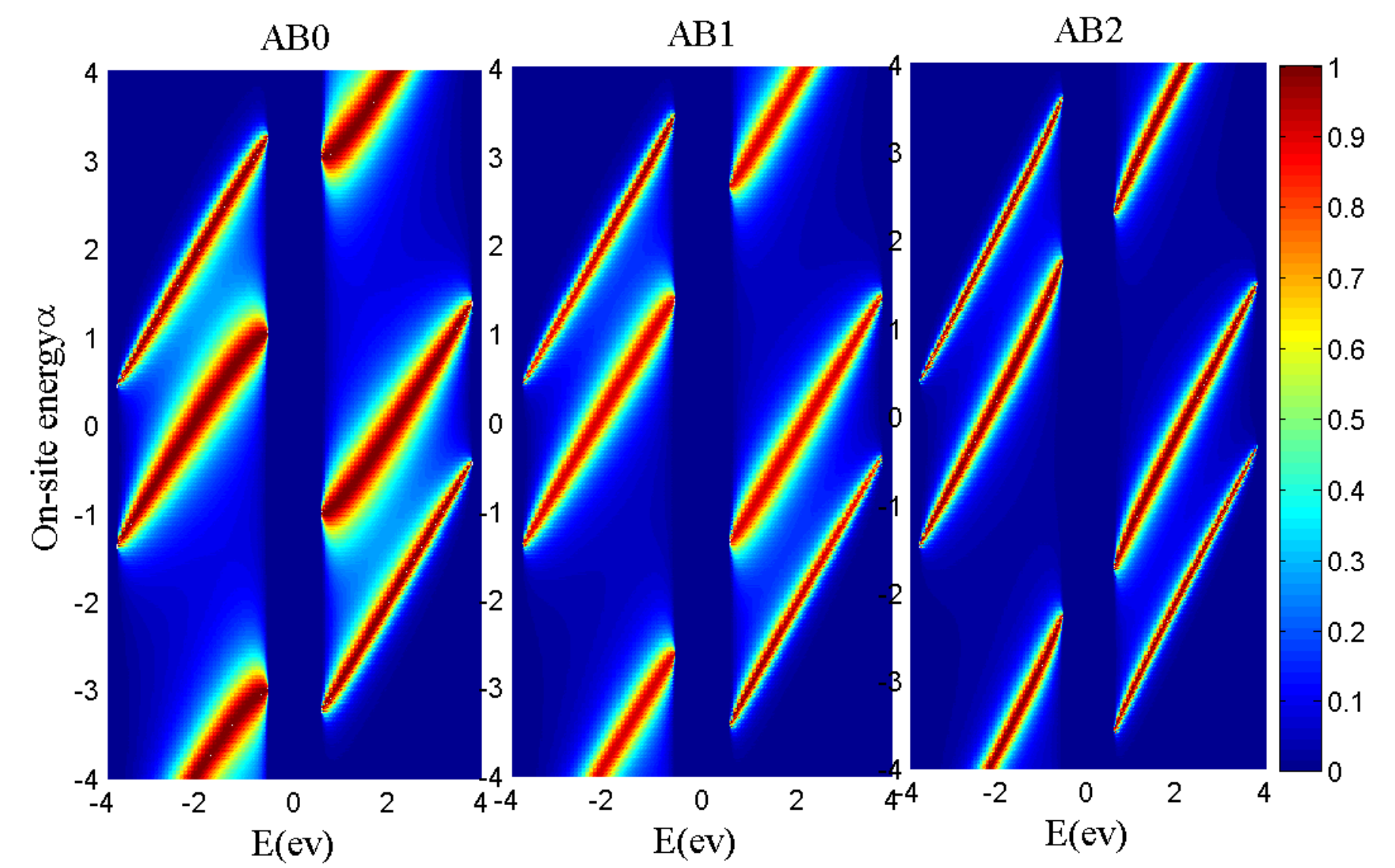}
 \includegraphics[width=0.85\columnwidth]{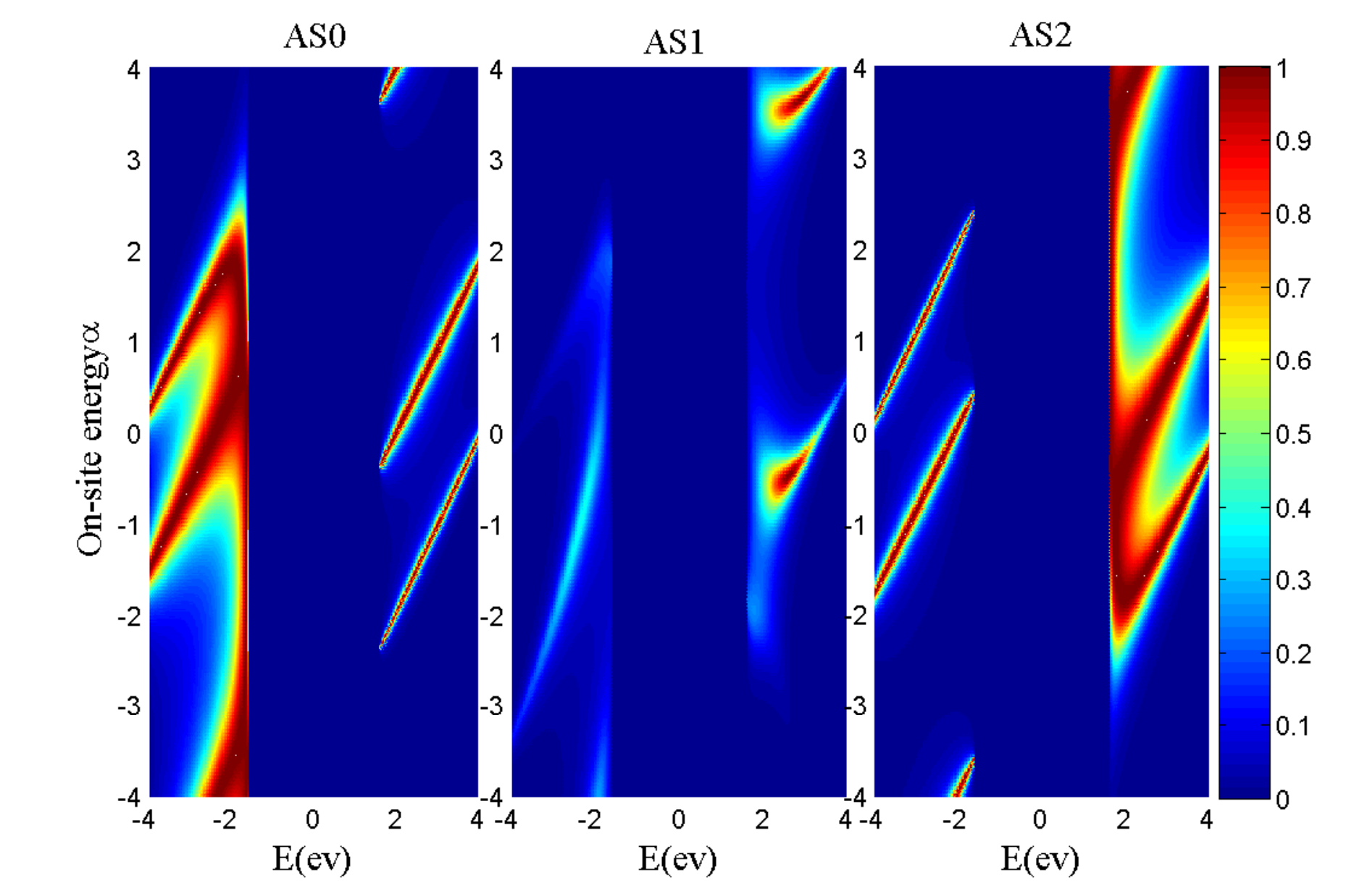}
 \includegraphics[width=0.95\columnwidth]{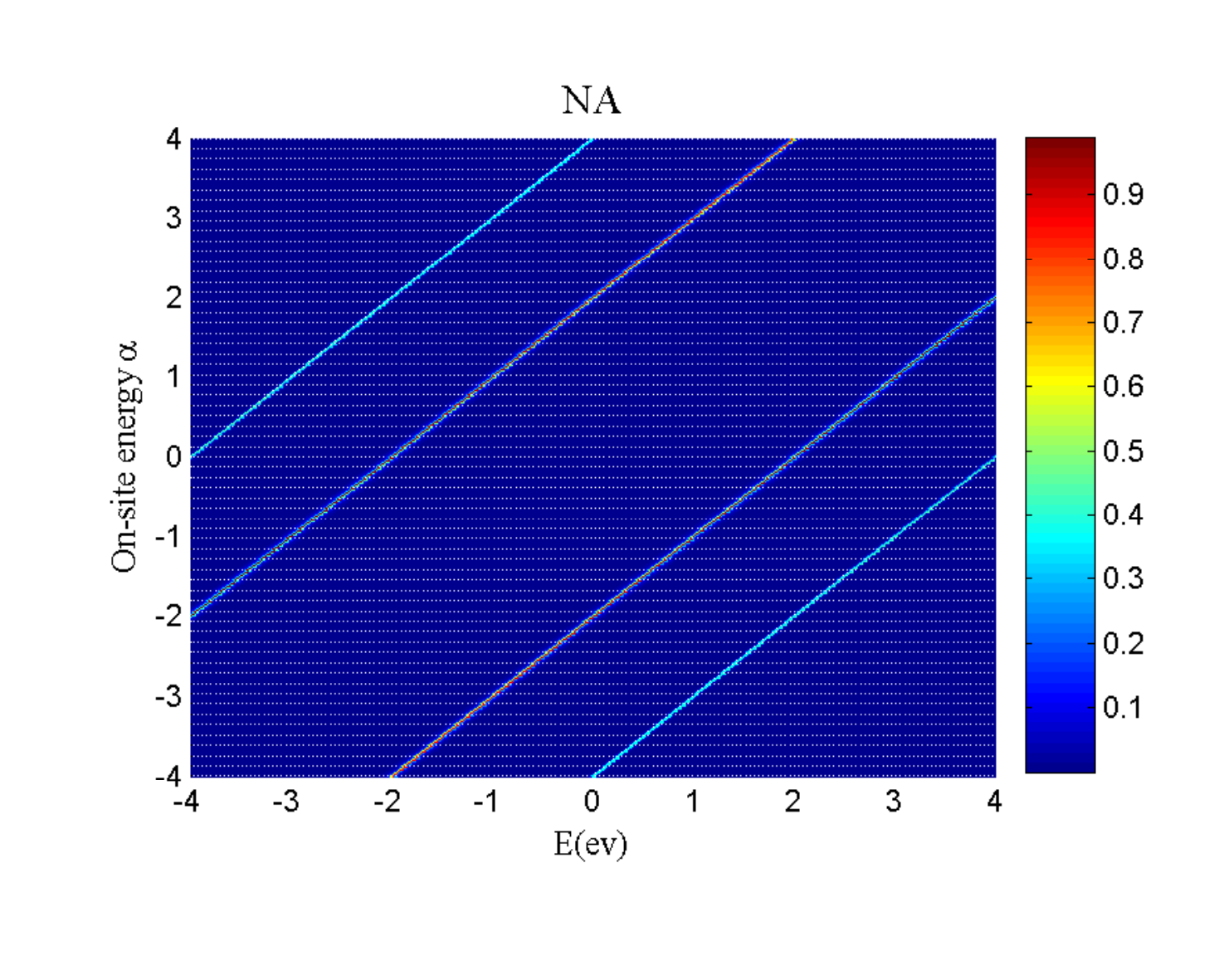}
 \caption{(Color online) Transmissions function of junction with two silicon (top panel),  titanium dioxide (middle panel) and gold (bottom panel) electrodes as a function of on-site energy of benzene molecule $\alpha$ and injected electron energy $E$.}
\label{fig7-1}
\end{figure}
\begin{figure}
 \includegraphics[width=0.75\columnwidth]{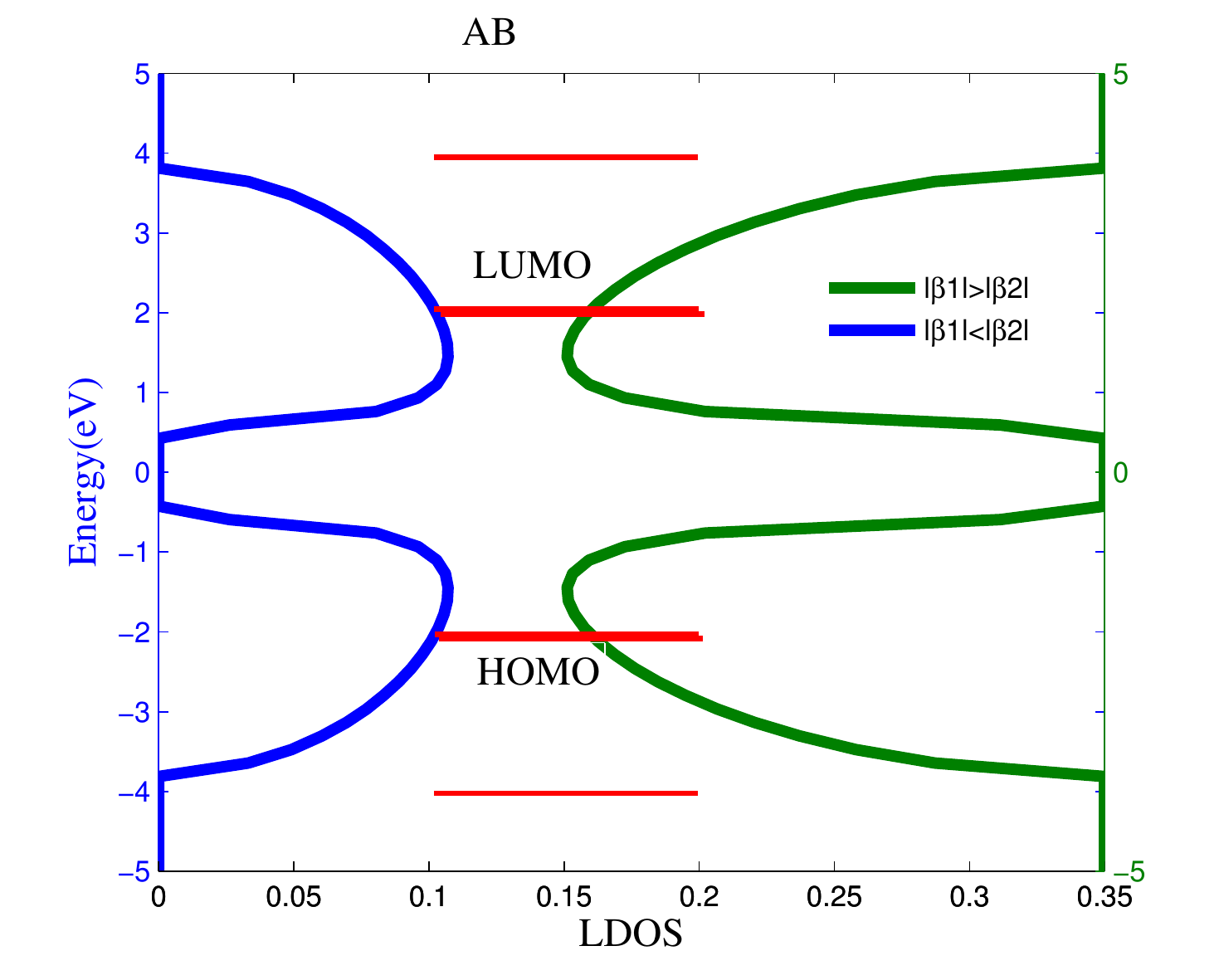}
 \includegraphics[width=0.75\columnwidth]{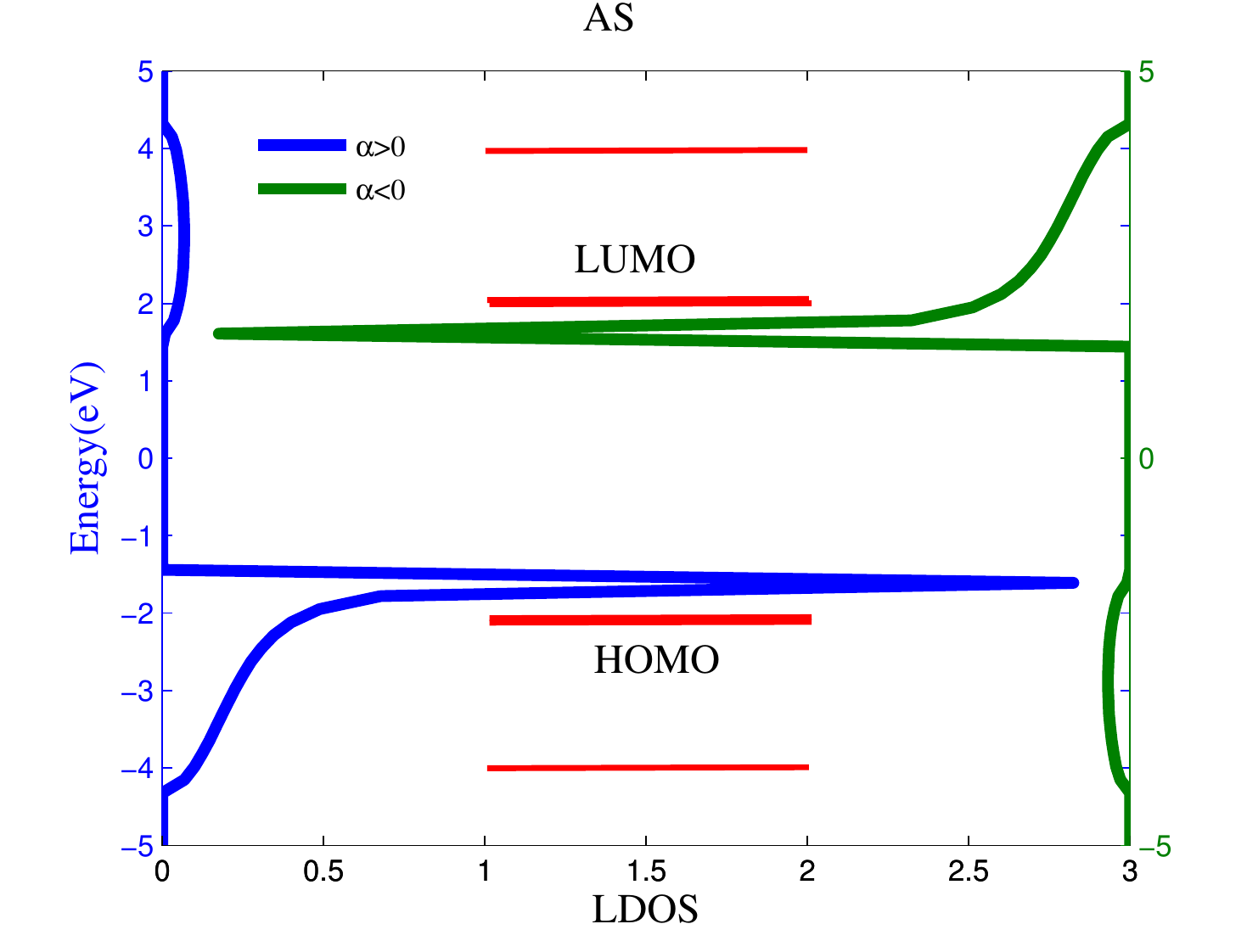}
 \includegraphics[width=0.75\columnwidth]{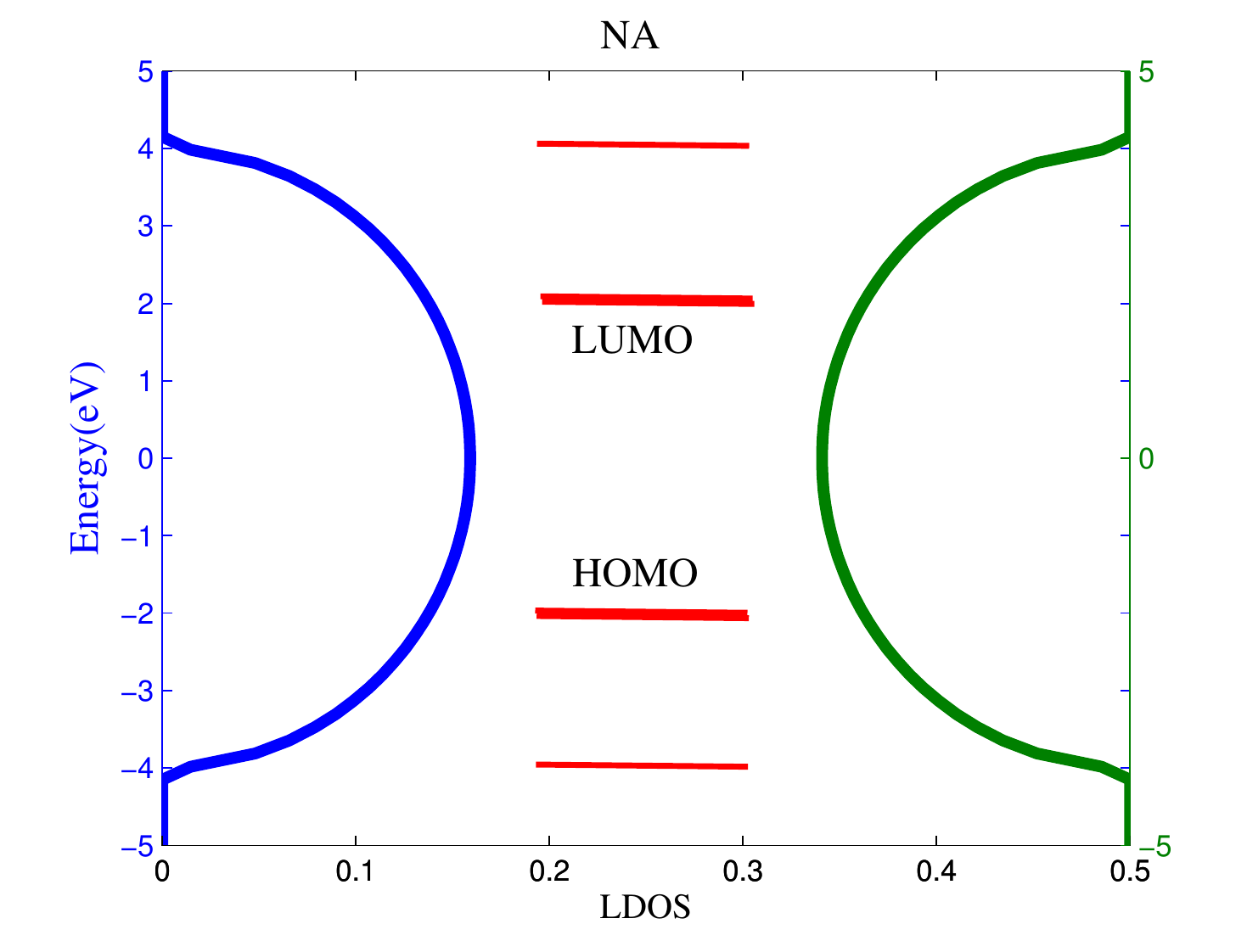}
 \caption{(Color online) Local density of states (LDOS) function for junction with two silicon (top panel),  titanium dioxide (middle panel) and gold (bottom panel) electrodes in the zero bias. The red lines are the energy levels of benzene.}
\label{fig7-2}
\end{figure}

\subsection{Semiconductor electrode}
\begin{figure*}
 \includegraphics[width=.68\columnwidth]{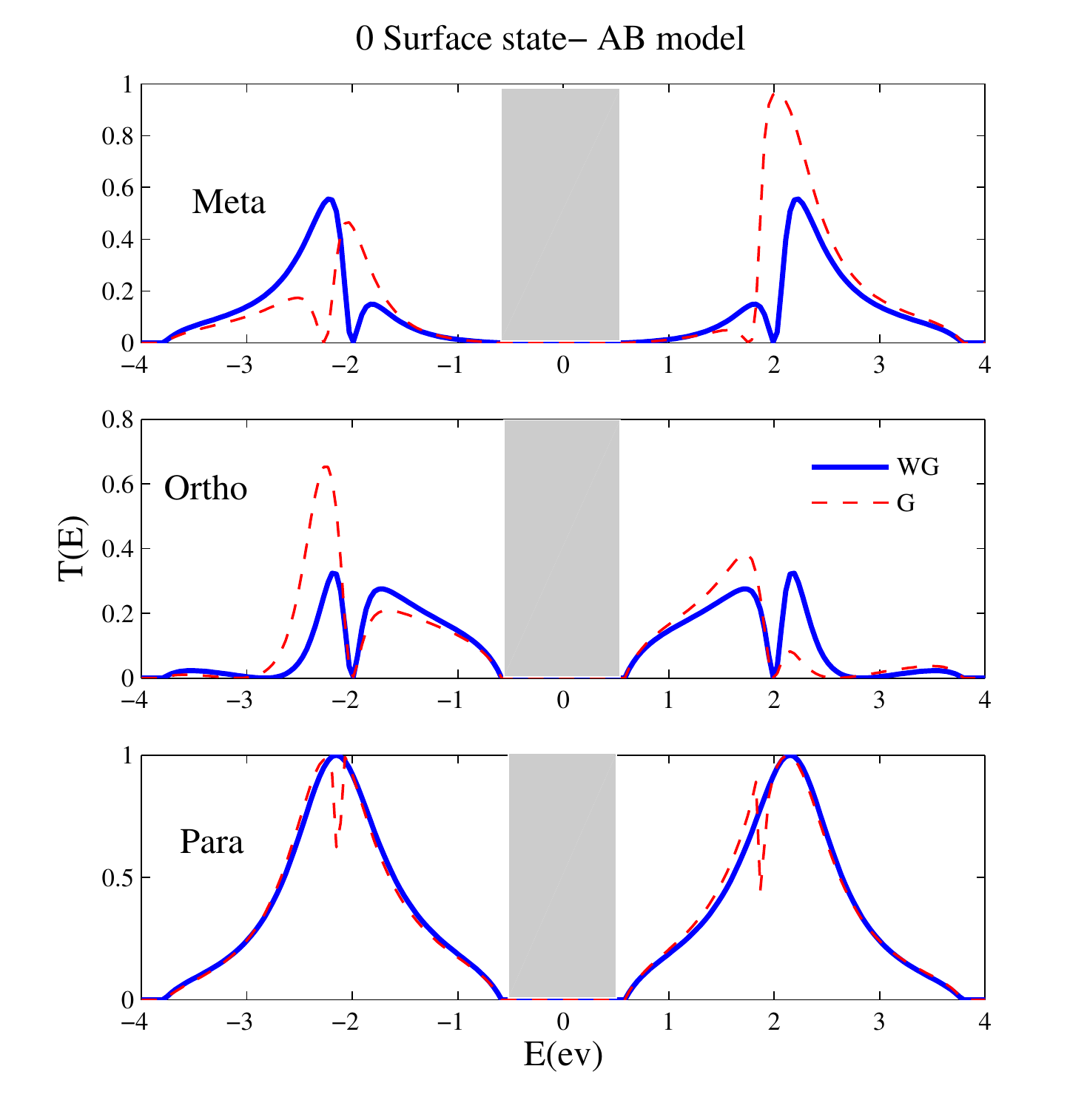}
 \includegraphics[width=.68\columnwidth]{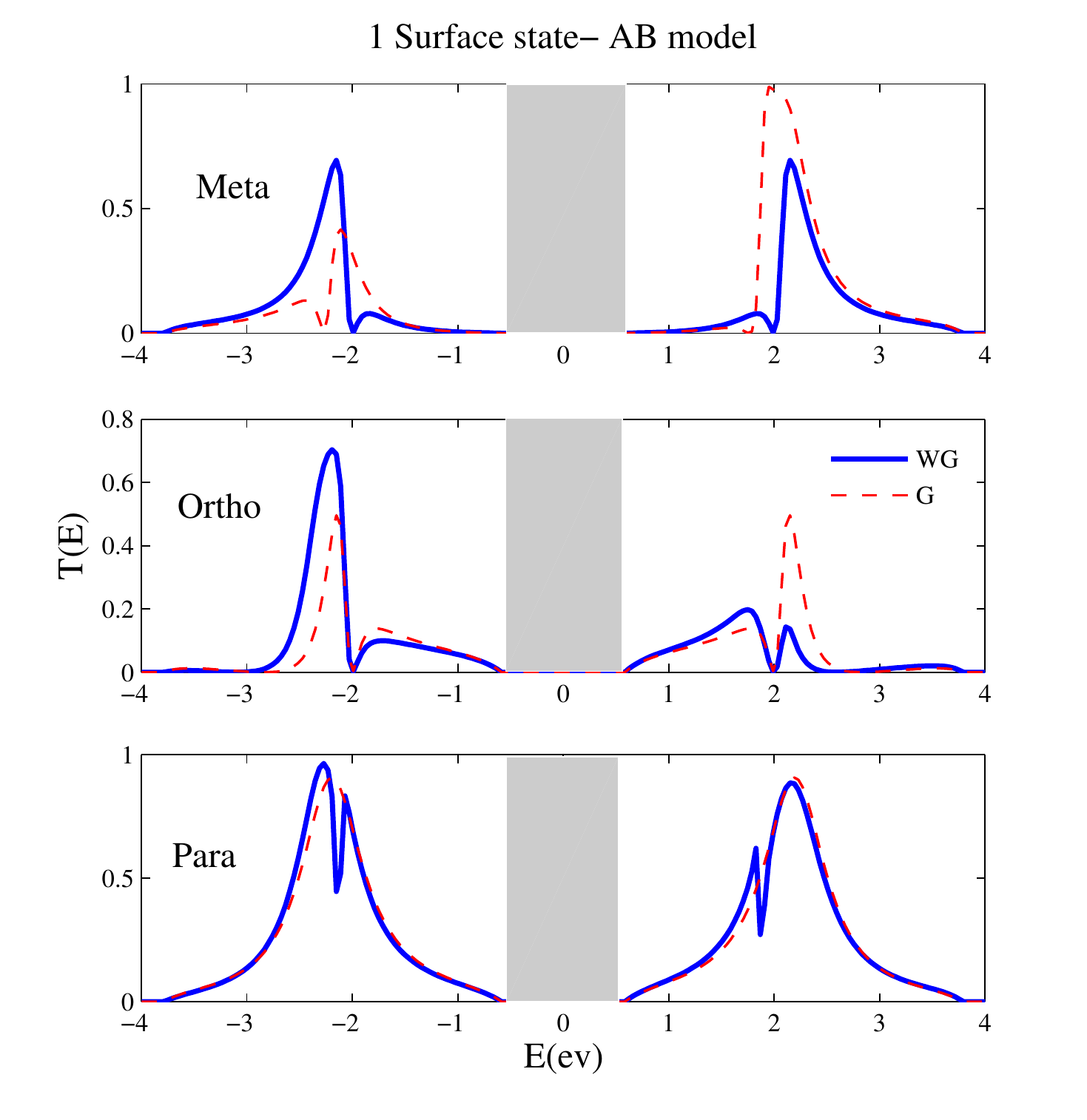}
 \includegraphics[width=.68\columnwidth]{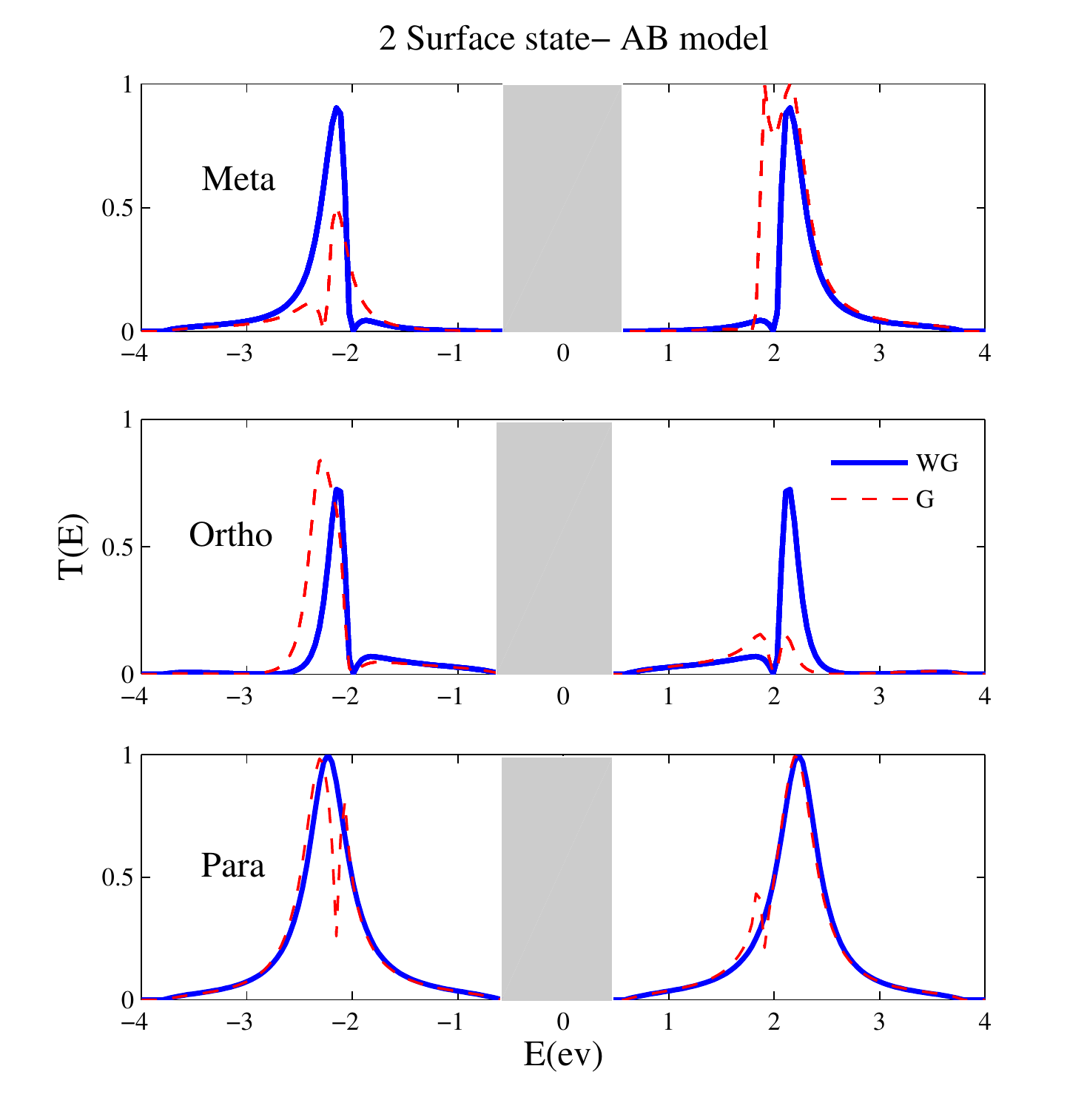}
 \caption{(Color online) Transmission function versus configurations in $0$, $1$ and $2$ surface states. The dotted (solid) lines corresponding to the gated (ungated) potential.} \label{fig7}
\end{figure*}

Having examined the metal/benzene/metal model in hand, we now consider semiconductor/benzene/semiconductor junction within the AS and AB models. We first take two silicon electrodes in the AB framework. The semiconductor band gap compared with the metal electrodes is the most noticeable feature in the transmission plots of Fig.(\ref{fig7},\ref{fig9}) as shown by the shadow region in each plot. In this region, the absence of states in the left electrode prevents electrons from injecting into the junction; also, there are no states for them to occupy once transmitted to the right electrode. A profound effect of semiconductor electrodes is the existence of  threshold bias voltage $V_t$ which is the minimum applied bias voltage needed to access states in either the valence or conduction band. 

Another important issue in the semiconductor-molecule junction, is the molecule-induced semiconductor surface states which can control charge transport across the junction\cite{a37,a38}. Authors in Ref. [33] have shown that surface states can significantly enhance transport through molecular energy levels and addressed this to the role of the real part of the self-energy, $\Lambda(E)$, which is non-negligible for semiconductor electrode. Fig.(\ref{fig7-1}) shows how transmission function changes for different electrodes. It is clear that semiconductor electrode generally broads the high transmission region and this is also sensitive how molecule bonds to the semiconductor surface states. For example, the AS model transmission spectrum shows less symmetry than the AB model which is the direct consequence of density of states. In Fig.(\ref{fig7-2}) we have depicted local density of states (LDOS) function for different electrodes. It is clear to see that LDOS of the AS model exhibits an asymmetry behavior versus energy. For detailed explain we refer the readers to the Ref. [33].
\begin{figure*}
 \includegraphics[width=.68\columnwidth]{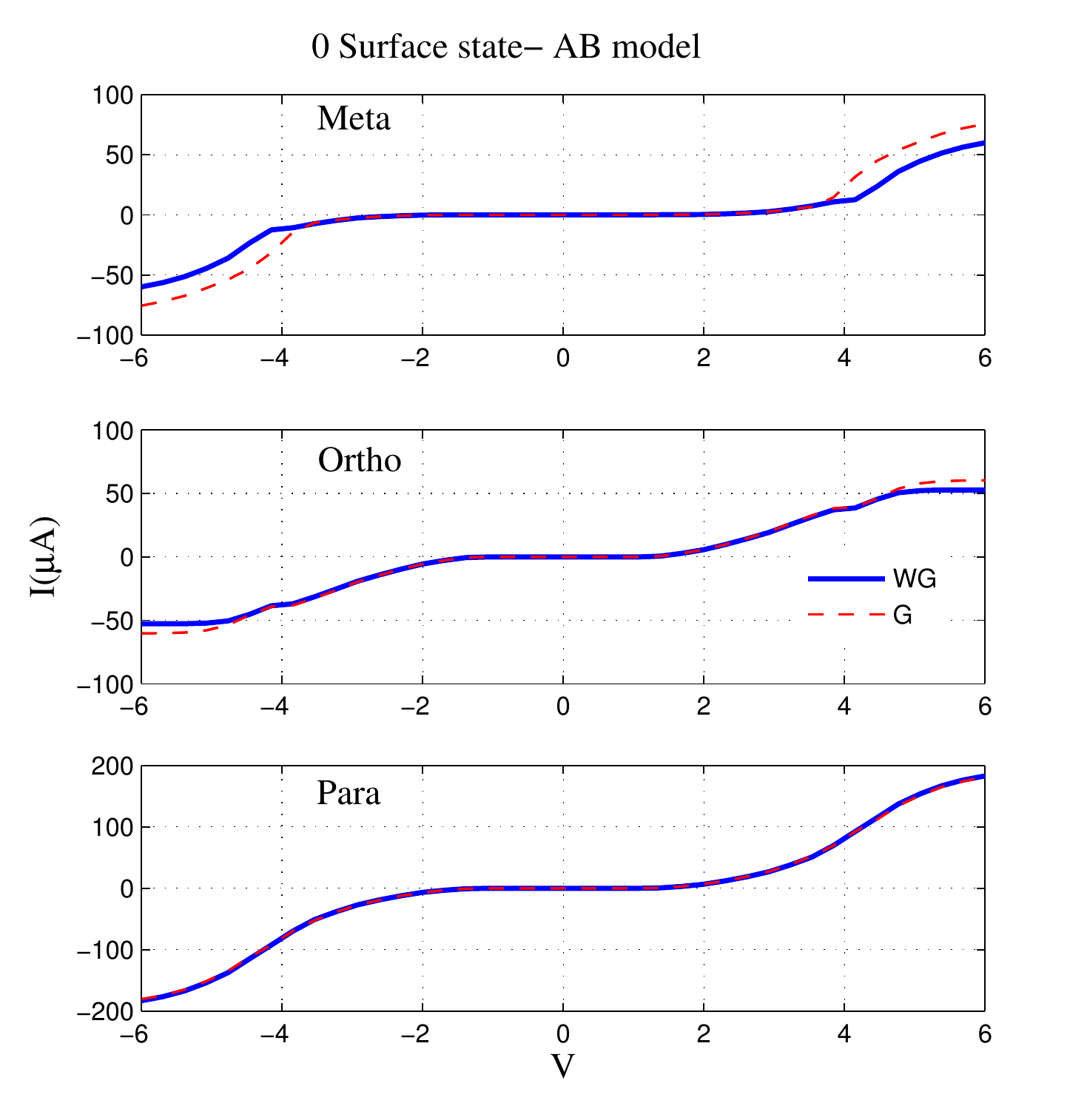}
 \includegraphics[width=.68\columnwidth]{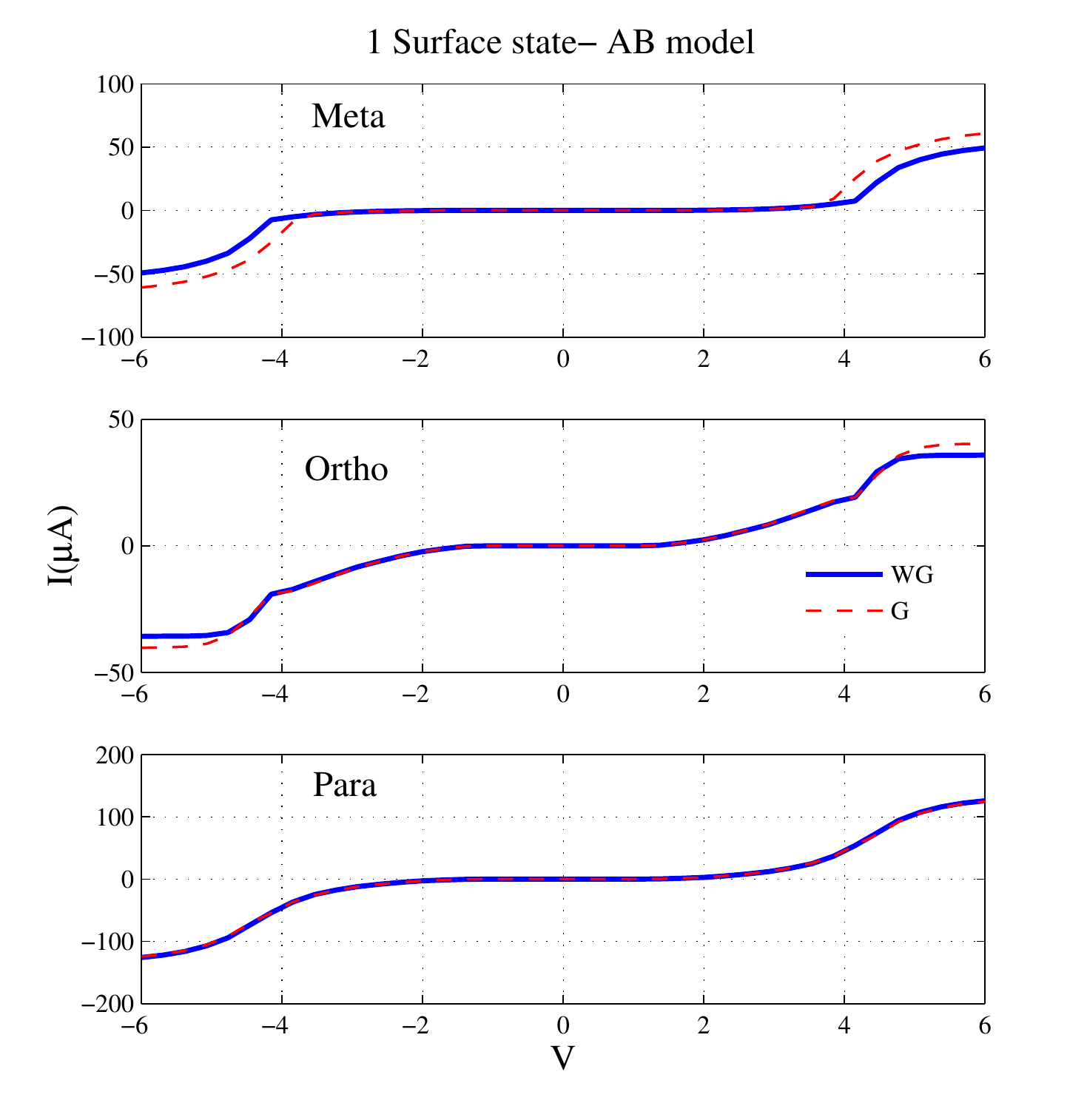}
 \includegraphics[width=.68\columnwidth]{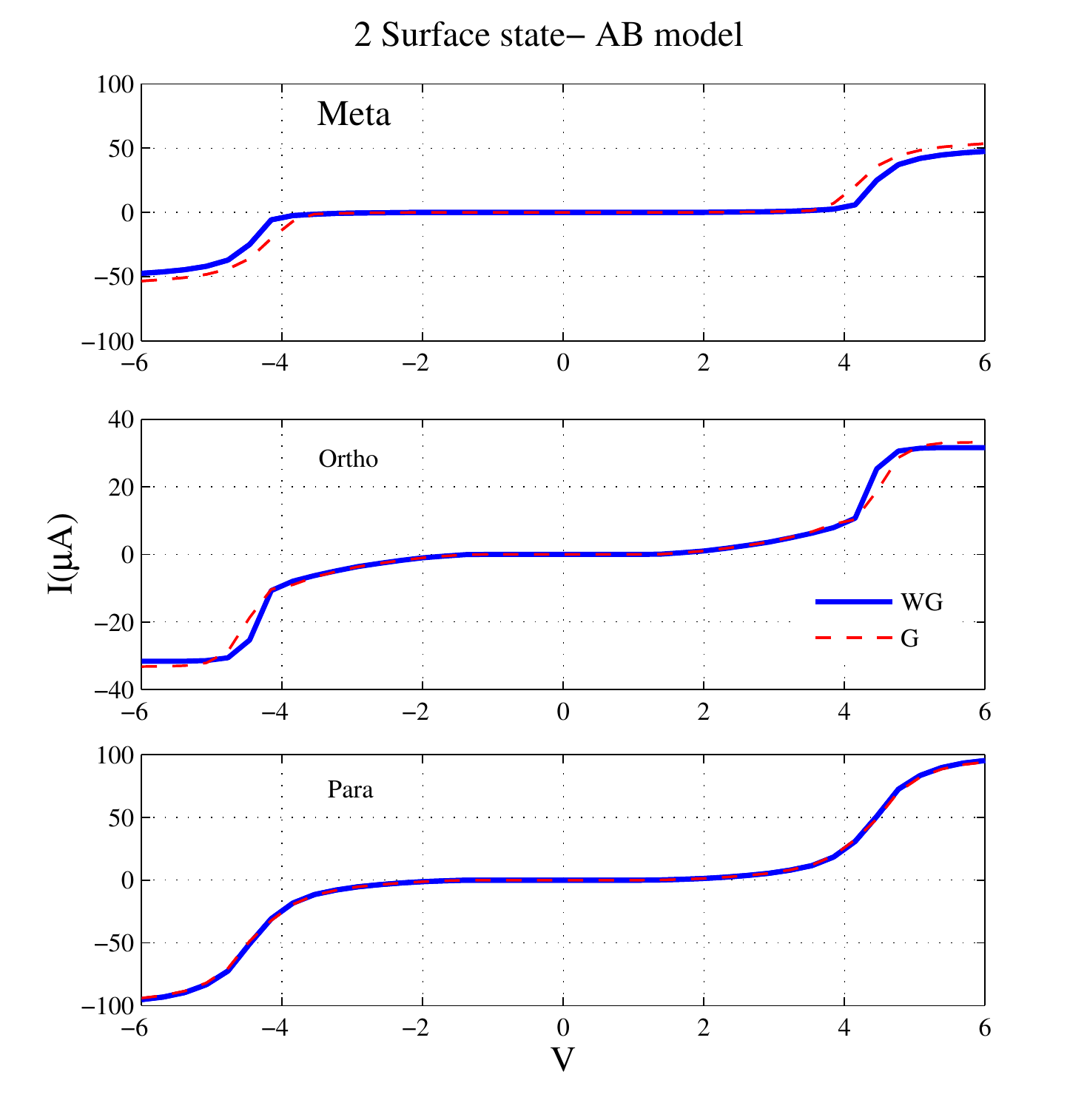}
 \caption{(Color online) The current-voltage (I-V) of the silicon$/$benzene$/$silicon junction for meta, ortho and para configurations in $0$ ,$1$ and $2$ surface states. The dotted (solid) lines corresponding to the gated (ungated) potential.}
\label{fig8}
\end{figure*}
\par
\textbf{\emph{Alternating bond model(AB)}:}  In Fig.(\ref{fig7}), we have plotted logarithmic scale of transmission function versus energy for three meta, ortho and para configurations, by considering $0$, $1$ and $2$ surface states. As it can be seen from this figure transmission function shows equal anti-resonance points in comparison with the metal electrode. These points had been predicted by the graphical approach. Note, applying a gate potential only makes a tiny shift on the transmission probability and the anti-resonance points.  However, in the ortho arrangement there is an anti-resonance point which is not shifted. Indeed, in all the three surface states the first anti-resonance peck occurs at the same point in gated and ungated cases.  In the para configuration for the all three surface states, in the absence of gate potential there is not any anti-resonance nodes, which can be addressed to the constructive superposition  of the electron waves, but immediately after applying the gate potential its transmission shows an anti-resonance behaviour (same as metal electrode), particularly in the $1$ surface state (dotted lines in the bottom panel of the three plots of Fig.(\ref{fig7}). In Fig.(\ref{fig8}), we have presented current as a function of voltage for meta, ortho and para configurations and the three surface states. All the three current-voltage characteristics show stair-like feature. The notable effect which should be mentioned is that the silicon electrodes cause different threshold voltage $V_t$ in comparison with gold electrodes in all the three meta, ortho and para configurations. Actually,  in the presence of silicon electrode threshold voltage $V_t$ occurs at lower bias. This effect is more profound in the ortho case. Furthermore, in the meta configuration for all the three surface states, current shows lower threshold voltage $V_t$ in the gated case than the ungated one. By comparing all the three surface states, maximum magnitude of current amplitude is seen in the para configuration for the $0$ surface state and the minimum one is seen in the ortho configuration for the $2$ surface state.
 \begin{figure*}
 \includegraphics[width=.68\columnwidth]{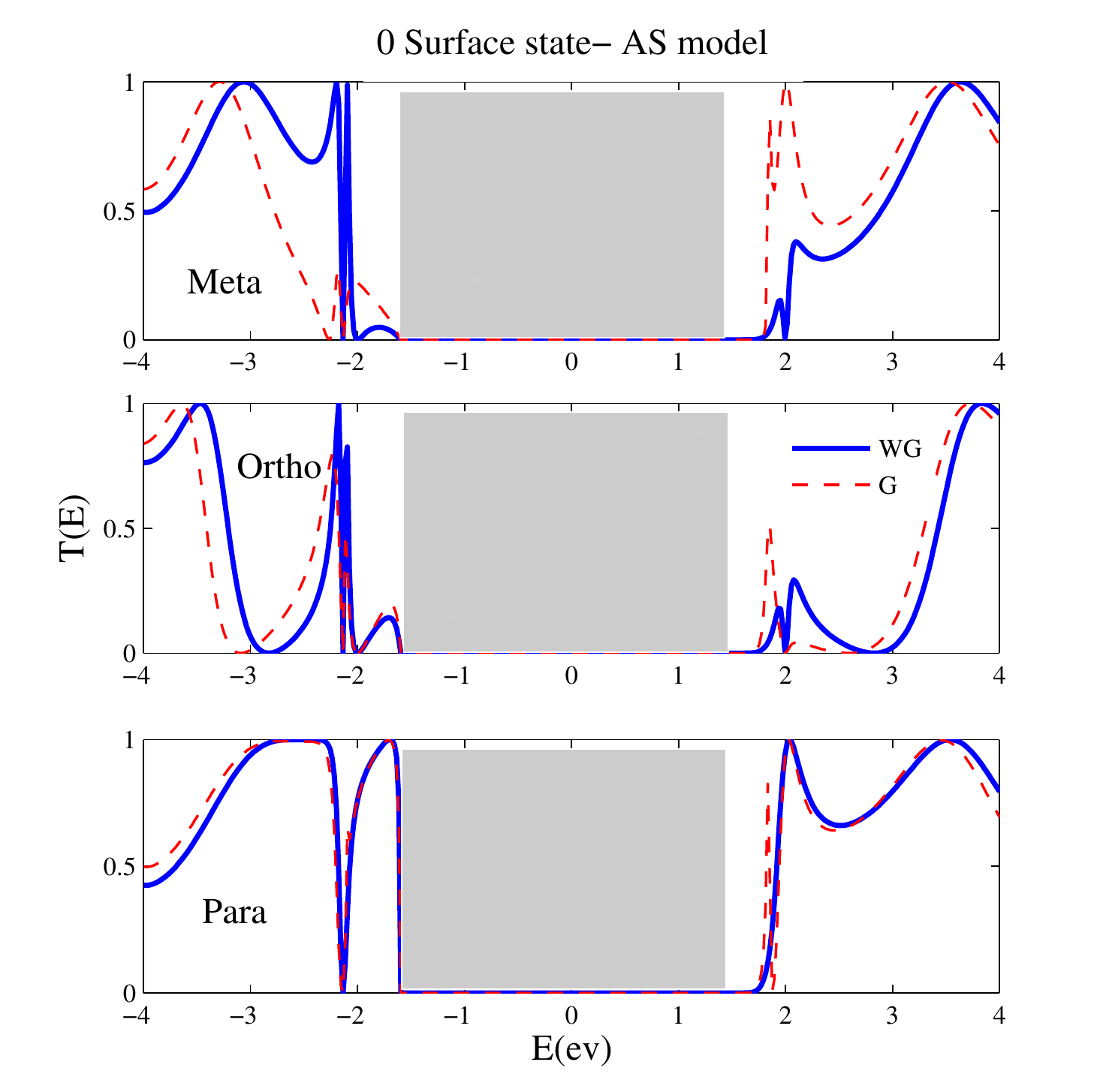}
 \includegraphics[width=.68\columnwidth]{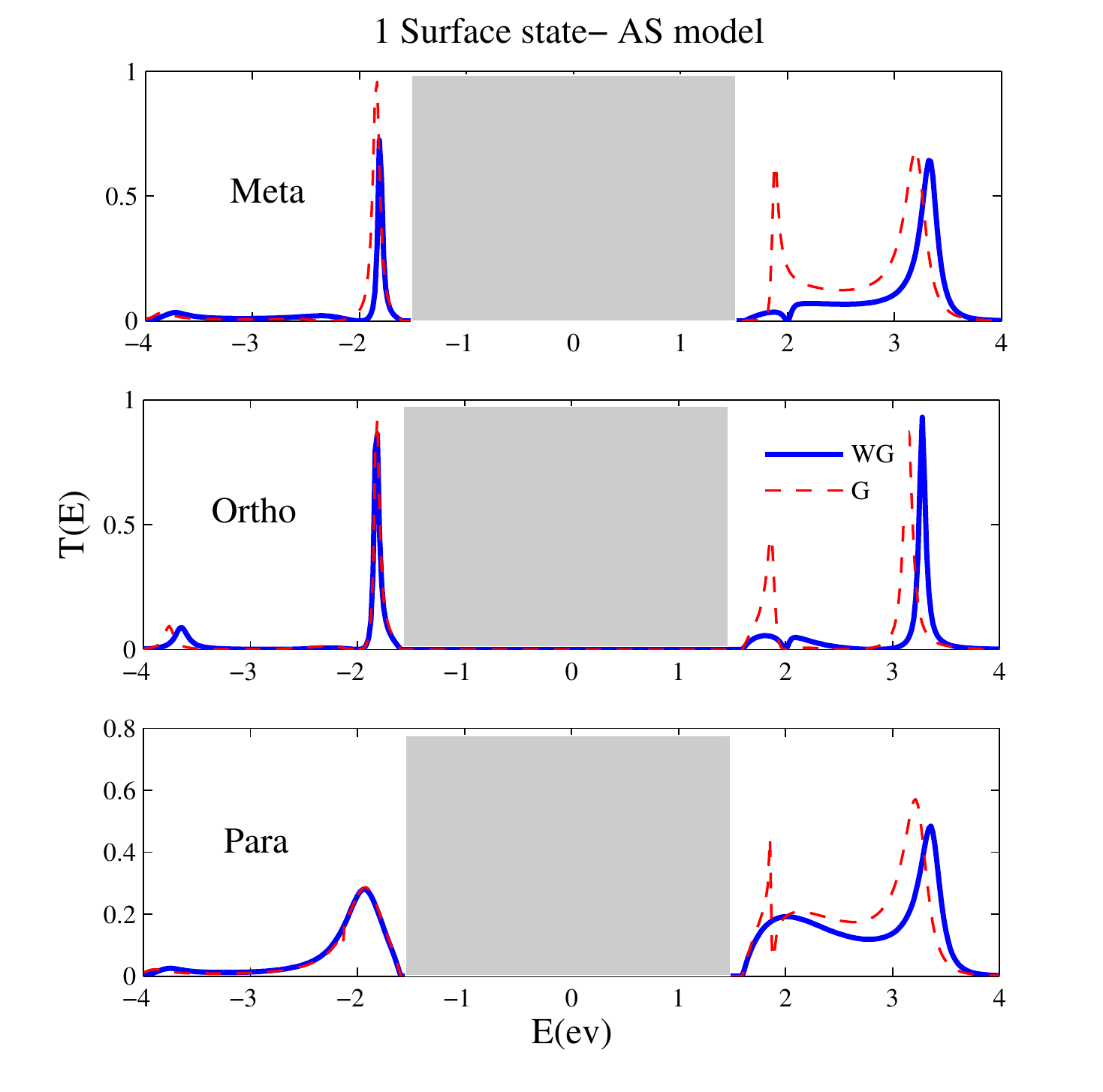}
 \includegraphics[width=.68\columnwidth]{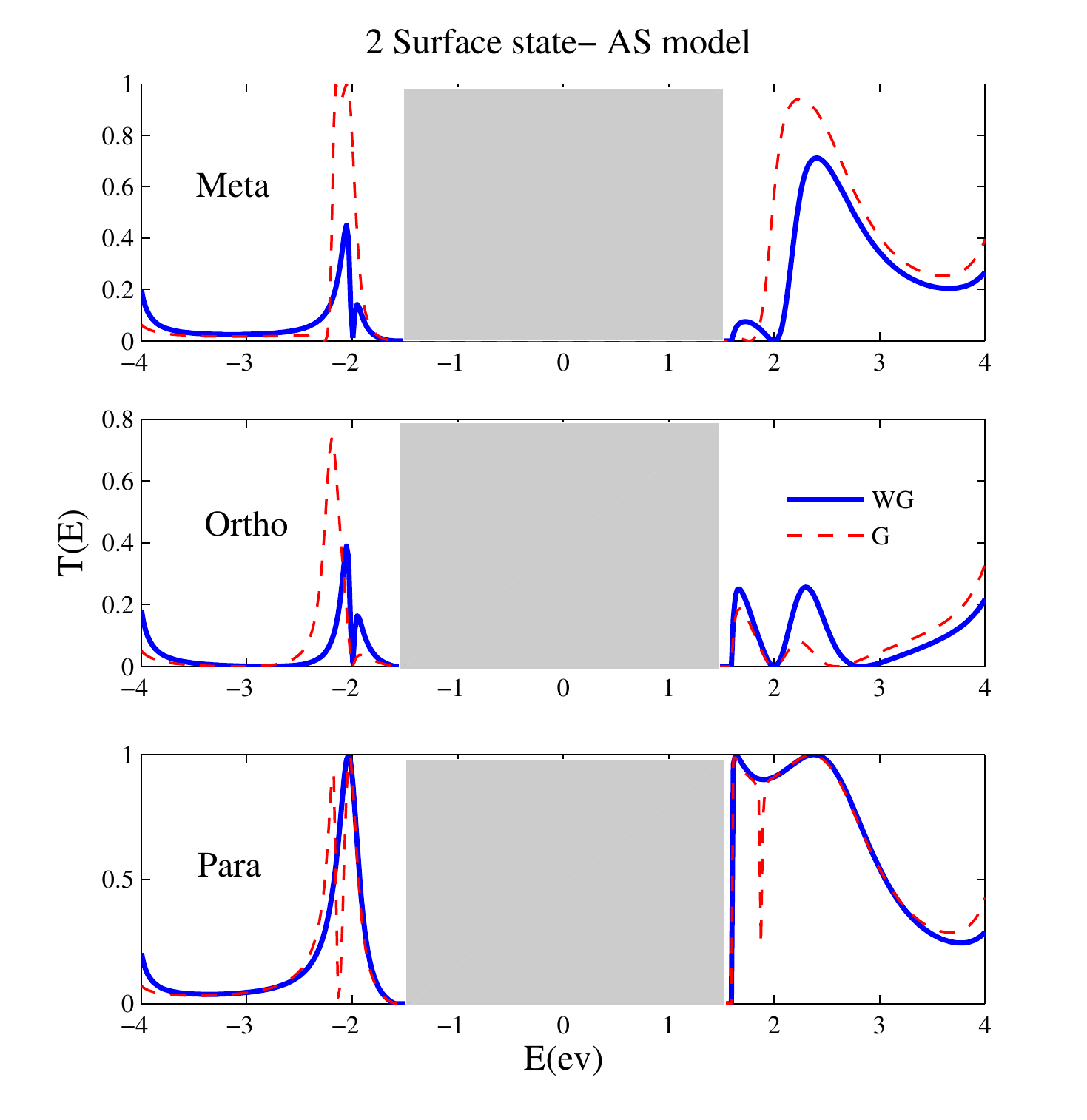}
 \caption{(Color online) Transmission function versus energy of the  titanium-dioxide/benzene/titanium-dioxide junction for meta, ortho and para configurations in $0$, $1$ and $2$ surface states. The dotted (solid) lines corresponding to the gated (ungated) potential.}
\label{fig9}
\end{figure*}
 \begin{figure*}
 \includegraphics[width=.68\columnwidth]{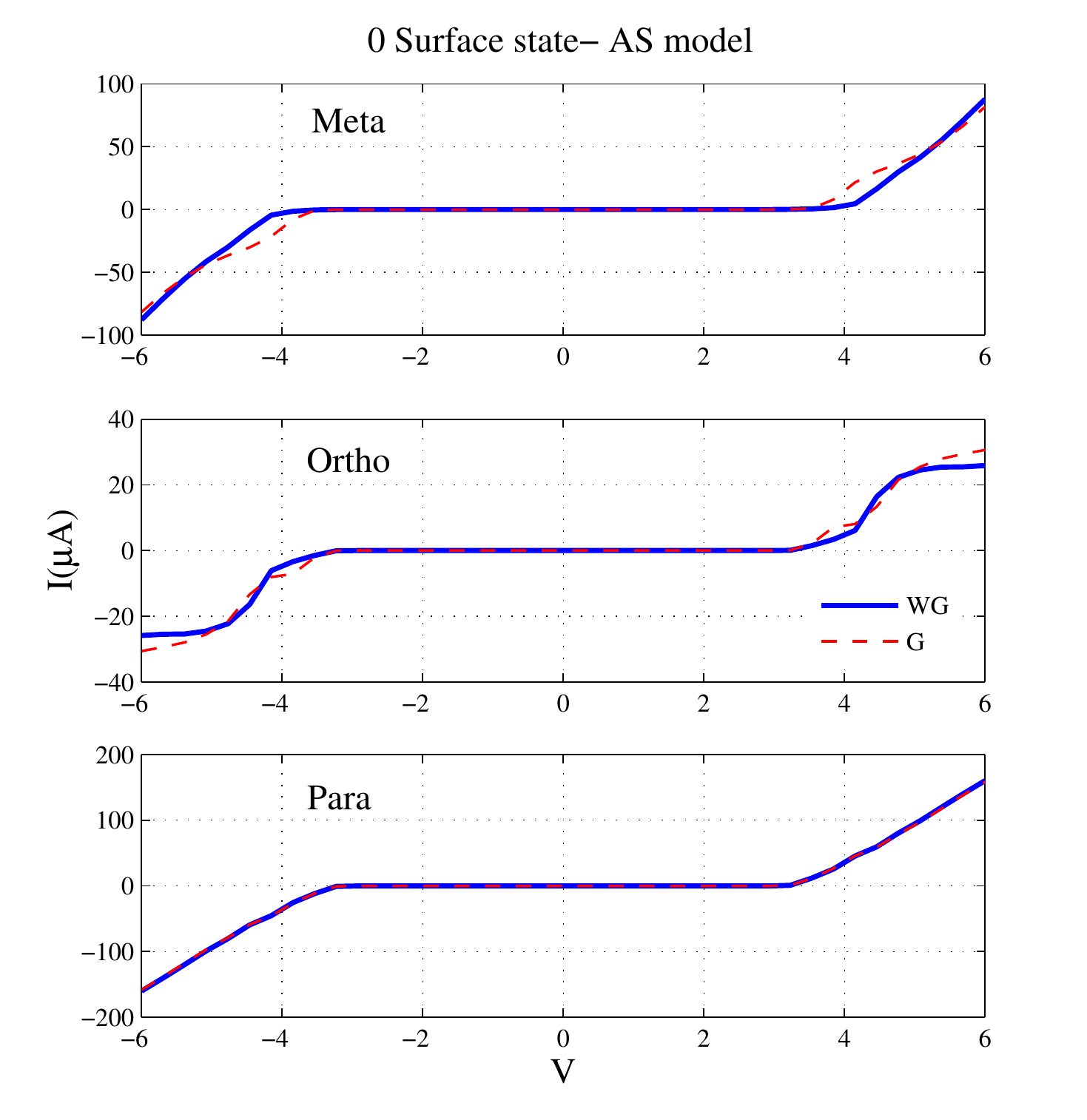}
 \includegraphics[width=.68\columnwidth]{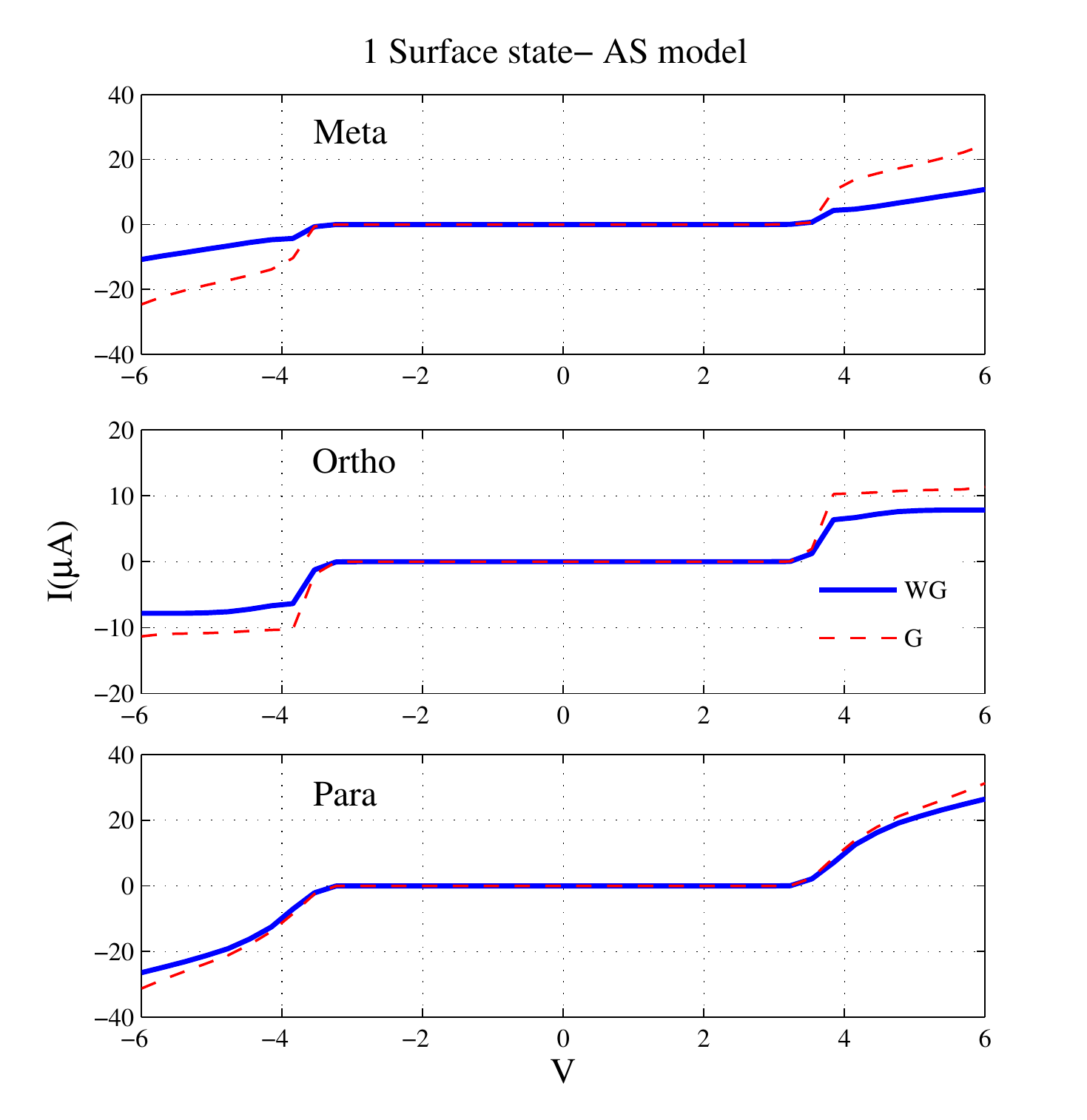}
 \includegraphics[width=.68\columnwidth]{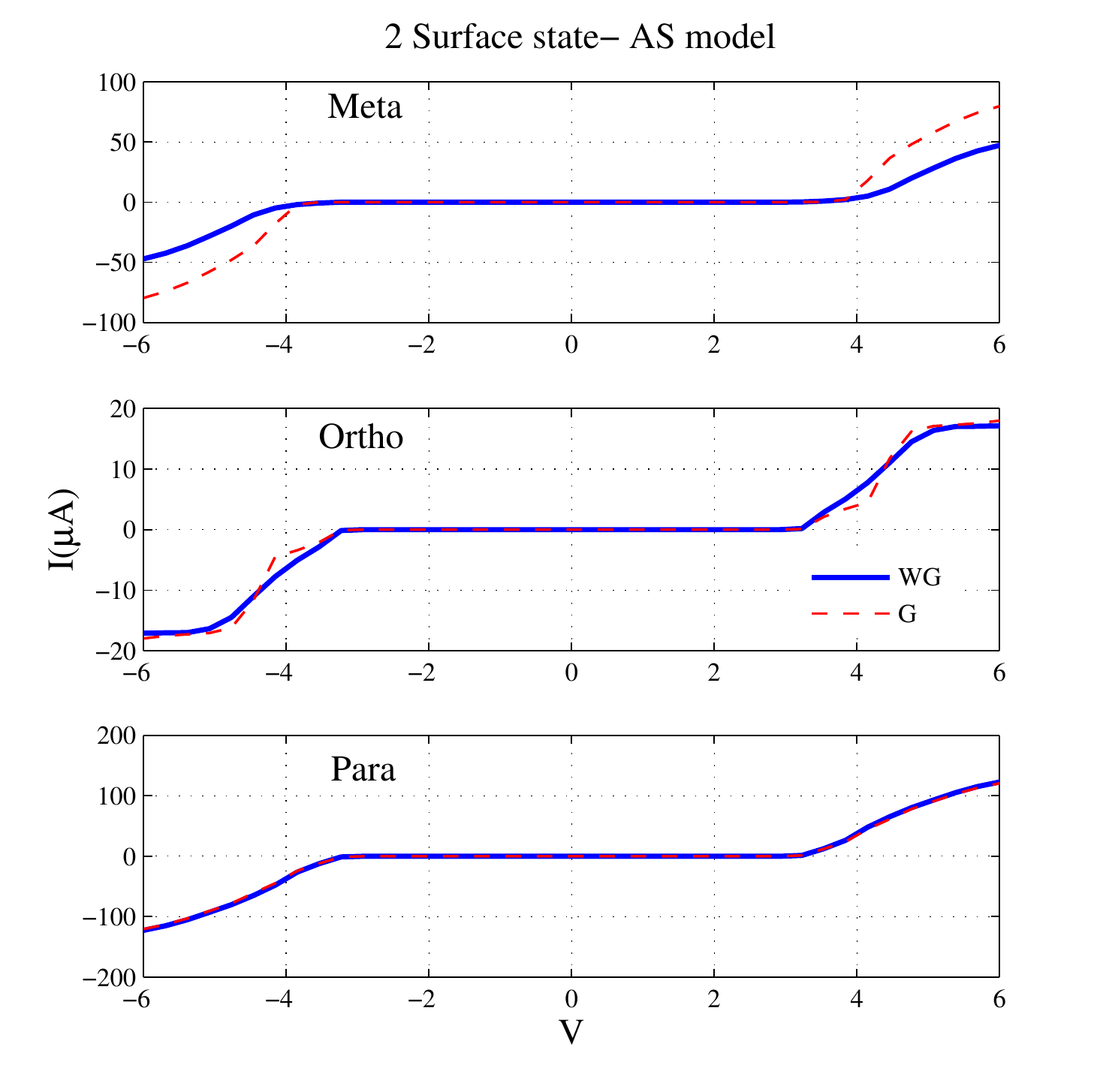}
 \caption{(Color online) The current-voltage (I-V) of the titanium-dioxide/benzene/titanium-dioxide junction for meta, ortho and para configurations in $0$ ,$1$ and $2$ surface states. The dotted (solid) lines corresponding to the gated (ungated) potential.}
\label{fig10}
\end{figure*}
\par
\textbf{\emph{Alternating site model(AS)}:} We now consider titanium dioxide ($TiO_2$) electrodes in the AS framework. In Fig.(\ref{fig9}), we have plotted  logarithmic scale of transmission function versus energy  for three meta, ortho and para configurations in the presence and absence of gate potential which  are shown in dotted and solid lines, respectively. 

For meta and ortho configurations in the top and middle panels of Fig.(\ref{fig9}), for all the three surface states, anti-resonance states happen in certain values of eigenenergies, due to the destructive quantum interference. The application of the gate potential produces a small change in the energy eigenvalues with respect to the ungated molecule and shifts the anti-resonance peaks. In Fig.(\ref{fig10}), we have illustrated the current as a function of applied voltage for the three meta, ortho and para configurations and the three surface states. For all the three meta, ortho and para configurations in the $0$ surface state applying gate potential don't have any significant effects on the current profile. However, in the $1$ surface state the applied gate one shows the most different behaviour respect to the ungated case. The minimum of current amplitude is also seen in the $1$ surface state. By comparing all the three meta, ortho and para configurations in the different surface states, one can easily find that the current has almost the same threshold voltage in the gated and ungated cases.
\section{CONCLUSION}\label{sec4 }
We numerically investigated the coherent electron transport through a benzene molecule sandwiched between  metal or semiconductor electrodes. Two generalized tight-binding models, alternating site (AS) and alternating bond (AB) models, have been considered to represent the semiconductor leads. The well known Newns-Anderson model has been also used to mimic the metal leads as a comparison benchmark. Due to symmetry breaking in the semiconductor leads surface states are formed. In this regard three kinds of surface states labeled as $0, 1$  and $2$ have been considered. The main goal  of this report is to study the effect of semiconductor electrode on the quantum interference characteristics of the benzene molecule. To this end, we have assumed three different terminal contact configurations (labeled as meta, ortho and para) and also a gate potential in order to verify the presence of quantum interference. Using the generalized Green's function technique based on the tight-binding model and the Landauer-B\"{u}ttiker theory, we have calculated the transmission probability function and the current-voltage characteristics. 

The first remarkable feature is the existence of a bias voltage threshold $V_t$ in the transmission probability function and the current-voltage figures. This threshold is linked to the hl{semiconductor} band gap and it is the minimum applied bias voltage needed to access the states in either the valence or conduction band. The second observed behavior is that the profile of the transmission probability function semiconductor electrodes shows an irregular pattern, in contrast to  the metal electrode. This effect can be explained by  molecular-level shifting imposed by adsorption to semiconductors and are very sensitive to how the molecule bonds to the semiconductor surface.  Nevertheless, in the integrated quantities, such as the current-voltage characteristics, no such behaviour is observed. The third noticeable feature is that the current magnitudes for the semiconductor/benzene/semiconductor systems are higher than its metal counterpart which is linked to the broadening of transmission probability function.

The gate potential can also create some anti-resonance peaks in the transmission probability and has a profound effect on the current magnitude in the meta configuration in both semiconductor and metal electrode cases. Moreover, we have explained how the quantum interference (QI) can create zeroes in the transmission function of the benzene molecule for all three electrode configurations using  simple graphical schemas involving solely the topology of the molecule. We hope the results open up new possibilities for the design of single-molecule devices sandwiched between semiconductor leads based on quantum interference effects, for instance, switching devices that operate by combining destructive and constructive molecular structures.
\\
\appendix
\section{Calculation of self-energy $\Sigma_{KD}(E)$} \label{App:AppendixA}
In order to have a self-contained and pedagogical manuscript, we have tried to go through the details of Eq.(\ref{e8},\ref{e9}). In this regard, we adapt the procedure used in Ref.[\cite{a33}]  to find the solution step by step. We start by Eq.(\ref{e7}), $\Sigma_{KD}(E)=\gamma_c^\ast g_{KD}\gamma_c=\Lambda_{KD}(E)-\frac{i}{2}\Gamma_{KD}(E)$,
 where $g_{KD}(E)=\left((E)-H_{KD}\right)^{-1}$ is the surface green function of the left/right electrode and $H_{KD}$ is their Hamiltonian
\begin{equation}
   H_{KD}=
  \left[ {\begin{array}{cccccc}
   \alpha & \beta_1 & 0 & 0 & 0 &  \cdots \\
   \beta_1 & -\alpha & \beta_2 & 0 & 0 &  \cdots \\
   0 & \beta_2 & \alpha & \beta_1 & 0 &  \cdots \\
   0 & 0 & \beta_1 & -\alpha & \beta_2 & \cdots\\
   0 & 0 & 0 & \beta_2& \alpha &  \cdots\\
   \vdots & \vdots & \vdots & \vdots&\vdots  & \ddots\\
  \end{array} } \right]
\end{equation}
In general the calculation of the Green's function of an operator $H$ requires the preliminary diagonalization of $H$. However, for tridiagonal operators the calculation is straightforward. For operator $H$ which has the tridiagonal matrix form, the operator$E-H$ is also tridiagonal form. So one can write the Green's function diagonal
matrix element as continued fraction. For our purpose which we are interested in surface Green's function one can achieve the following expression
\begin{equation}
  g_{KD}(E) =  \cfrac{1}{E-\alpha
       - \cfrac{\beta_1^2}{E+\alpha
       - \cfrac{\beta_2^2}{E-\alpha
       + \cfrac{\beta_1^2}{\cdots} } } }
\end{equation}
\par
To perform the continued fraction sum, we recast the above equation as
\begin{equation}
  g_{KD}(E) =  \cfrac{1}{E-\alpha
       - \cfrac{\beta_1^2}{E+\alpha-\beta_2^2 g_{KD}(E)  } }
\end{equation}
after a simple calculation one can find Eq.(\ref{e9}). Now substitution this solution into the self-energy it is easy to have the following expression
\begin{widetext}
\begin{eqnarray}
\frac{\Sigma_{KD}(E)}{\gamma^2}&=&\Lambda_{KD}(E)-\frac{i}{2}\Gamma_{KD}(E)\nonumber\\
                               &=&\frac{E^2-\alpha^2-\beta_1^2+\beta_2^2\pm \sqrt{\big[E^2-\alpha^2-(\beta_1-\beta_2)^2\big] \big[E^2-\alpha^2-(\beta_1+\beta_2)^2\big]}}{2\beta_2^2(E-\alpha)}
\end{eqnarray}
\end{widetext}
having this equation, it is simple to find $\Sigma_{NA}(E)$, $\Sigma_{AB}(E)$ and $\Sigma_{AS}(E)$ equations. Here we For the NS model
by setting $\alpha=0$ and $\beta_1=\beta_2=\beta$ we have $\frac{\Sigma_{NA}(E)}{\gamma^2}=\frac{E\pm\sqrt{E^2-4\beta^2}}{2\beta^2}$.
It is easy to see that $\Sigma_{NA}$ is complex for $E^2-4\beta^2<0 (|E|<2|\beta|)$, but for $E<-2|\beta|$ and $E>2|\beta|$ is real
and up to a sign it shows $\Sigma_{NA}(|E|\rightarrow\infty)\rightarrow 0$. The real and complex behaviour of $\Sigma_{NA}(E)$ can be referred to
shifts and broadening of molecular energy levels. So imaginary part of self-energy $Im\left(\frac{\Sigma_{NA}(E)}{\gamma^2}\right)=\pm\frac{\sqrt{4\beta^2-E^2}}{2\beta^2}$
gives spectral density as $\Gamma_{NA}(E)=\frac{\gamma^2}{\beta^2}\sqrt(4\beta^2-E^2)$ and the real part of self-energy gives broadening as
$\frac{\Lambda_{NA}(E)}{\gamma^2}=\frac{E}{2\beta^2}+\Theta_{NA}\frac{\sqrt{E^2-4\beta^2}}{2\beta^2}$ where $\Theta_{NA}=\Theta(-2|\beta|-E)-\Theta(E-2|\beta|)$. The same procedure
holds for $\Sigma_{AB}(E)$ and $\Sigma_{AS}(E)$ equations and it is easy to ascertain.


\vspace{0.3cm}


\section*{References}
\begin{thebibliography}{10}
\bibitem{a1} G. Cuniberti, G. Fagas and K. Richter, Introducing Molecular Electronics., Lect. Notes Phys. {\bf 680} (Springer, NewYork, 2005).
\bibitem{a2} R.Baer and D. Neuhauser, J. Am. Chem. Soc. {\bf 124}, 4200(2001).
\bibitem{a3} R. Stadler, M. Forshaw and C. Joachim, Nanotechnology {\bf 14}, 138 (2003).
\bibitem{a4} R. Stadler, S. Ami, M. Forshaw and C. Joachim, Nanotechnology {\bf 15}, S115 (2004).
\bibitem{a5} R. Stadler, K. S. Thygesen and K.W. Jacobsen, Nanotechnology {\bf 16}, S155 (2005).
\bibitem{a6} T. A. Papadopoulos, I. M. Grace and C. J. Lambert, Phys. Rev. B {\bf 74}, 193306 (2006).
\bibitem{a7} D. M. Cardamone, C. A. Stafford and S. Mazumdar, Nano Lett. {\bf 6}, 2422 (2006).
\bibitem{a8} C. A. Stafford, D. M. Cardamone and S. Mazumdar, Nanotechnology {\bf 18}, 424014 (2007).
\bibitem{a9} S. H. Ke, W. Yang and H. U. Baranger, Nano Lett. {\bf 8}, 3257 (2008).
\bibitem{a10} F. Sols, M. Macucci, U. Ravaioli and K. Hess, Appl. Phys. Lett. {\bf 54}, 350 (1989).
\bibitem{a11} W. Porod, Z. Shao and C. S. Lent, Appl. Phys. Lett. {\bf 61}, 1350 (1992).
\bibitem{a12} W. Porod, Z. Shao and C. S. Lent, Phys. Rev. B {\bf 48}, 8495 (1993).
\bibitem{a13} P. Debray, O. E. Raichev, P. Vasilopoulos, M. Rahman, R. Perrin and W. C. Mitchell, Phys. Rev. B {\bf 61}, 10950 (2000).
\bibitem{a14}  P. Sautet and C. Joachim, Chem. Phys. Lett. {\bf 153}, 511 (1988).
\bibitem{a15} C. Patoux, C. Coudret, J. P. Launay, C. Joachim and A. Gourdon, Inorg. Chem. {\bf 36}, 5037 (1997).
\bibitem{a16}R. Stadler, Phys. Rev. B. {\bf 80}, 125401(2009).
\bibitem{a17} M. Magoga and C. Joachim, Pys. Rev. B. {\bf 56}, 4722 (1997).
\bibitem{a18} S. N. Yaliraki and M. Ratner, A. J. Chem. Phys. {\bf 109}, 5036 (1998).
\bibitem{a19} E. G. Emberly and G. Kirczenow, Phys. Rev. B. {\bf 58}, 10911 (1998).
\bibitem{a20} L. E. Hall, J. R. Reimers, N. S. Hush and K. Silverbrook, J. Chem. Phys. {\bf 112}, 1510 (2000).
\bibitem{a21} J. Hihath, C. R. Arroyo, G. Rubio-Bollinger, N. Tao, and N. Agraït, Nano Lett. {\bf 8}, 1673 (2008).
\bibitem{a22} J. S. Kristensen, D. J, Mowbray, K. S. Thygesen, K. W.Jacobsen, J. Phys.: Condens. Matter. {\bf 20}, 374101 (2008).
\bibitem{a23} W. Tian, S. Datta, S. Hong, R. Reifenberger, J. I. Henderson and C. P. Kubiak, J. Chem. Phys. {\bf 109}, 2874 (1998).
\bibitem{a24} R. M. Metzger, T. Xu and I. R. Peterson, J. Phys. Chem. B. {\bf 105}, 7280 (2001).
\bibitem{a25} D. R. Ward, N. J. Halas, J. W. Ciszek, J. M. Tour, Y. Wu, P. Nordlander and D. Natelson,  Nano Lett. {\bf 8}, 919 (2008).
\bibitem{a26} R. McCreery, J. Dieringer, A. O. Solak, B. Snyder, A. M. Nowak, W. R. McGovern and S. Duvall, J. Am. Chem. Soc. {\bf 125}, 10748 (2003).
\bibitem{a27} N. P.Guisinger,  M. E.Greene, R. Basu, A. S. Baluch and M. C. Hersam, Nano Lett. {\bf 4}, 55 (2004).
\bibitem{a28} N. P.Guisinger,  N. L. Yoder and M. C. Hersam, Proc. Natl. Acad. Sci. U.S.A.  {\bf 102}, 8838 (2005).
\bibitem{a29} P. G.Piva, G. A. DiLabio, J. L. Pitters, J. Zikovsky, M.  Rezeq, S. Dogel,  W. A. Hofer and R. A. Wolkow,  Nature {\bf 435}, 658 (2005).
\bibitem{a31} T. Rakshit, G. C. Liang, A. W. Ghosh and S. Datta, Nano Lett. {\bf 4}, 1803 (2004).
\bibitem{a32}  T. Rakshit, G. C. Liang, A. W. Ghosh, M. C. Hersam and S. Datta, Phys. ReV. B. {\bf 72}, 125305 (2005).
\bibitem{a322} D. Nozaki, S. M. Avdoshenko, H. Sevinçli, R. Gutierrez1 and G. Cuniberti, J. Phys.: Conf. Ser. {\bf 427}, 012013 (2013).
\bibitem{a33}  M. G. Reuter, T. Hansen, T. Seideman,and M. A. Ratner, Journal of Physical Chemistry A. {\bf 113}, 4665 (2009).
\bibitem{a34}  I. Tamm, Phys. Z. Sowjetunion {\bf 1}, 733 (1933).
\bibitem{a35}  W. Shockley, W. Phys. ReV. {\bf 56}, 317 (1939).
\bibitem{a36}  T. Markussen, R. Stadler and K. S. Thygesen, Nano Lett. {\bf 10}, 4260 (2010). T. Markussen, R. Stadler and K. S. Thygesen, Chem. Chem. Phys. {\bf 13}, 1431 (2011).
\bibitem{a37} N. L. Yoder, N. P. Guisinger, M. C. Hersam, R. Jorn, C. C. Kaun and T. Seideman, Phys. Rev. Lett. {\bf 97}, 187601 (2006).
\bibitem{a38}  L. H. Yu, N. G. Hackett, C. D. Zangmeister, C. A. Hacker, C. A. Richter and J. G. Kushmerick, J. Phys.: Condens. Matter {\bf 20}, 374114 (2008)

\end{thebibliography}
\end{document}